\documentclass[sn-mathphys-num]{sn-jnl}% Math and Physical Sciences Numbered Reference Style 

\usepackage{graphicx}%
\usepackage{multirow}%
\usepackage{amsmath,amssymb,amsfonts}%
\usepackage{amsthm}%
\usepackage{mathrsfs}%
\usepackage[title]{appendix}%
\usepackage{xcolor}%
\usepackage{textcomp}%
\usepackage{manyfoot}%
\usepackage{booktabs}%
\usepackage{algorithm}%
\usepackage{algorithmicx}%
\usepackage{algpseudocode}%
\usepackage{listings}%

\usepackage{bm} % for equation
\usepackage{nicefrac} % for equation
\usepackage{colortbl} % for tables
\usepackage{makecell} % for tables
\usepackage{tabularx} % for tables
\usepackage{multirow} % for tables
\usepackage{subfig} % Numbered and caption subfigures using \subfloat.
\usepackage{comment} % comments
\usepackage{subfloat} % for floats
\usepackage{fancyhdr} % for headers

\newcommand{\e}[1]{\times 10^{#1}}  % Powers of 10 notation
\begin{document}

\title[Article Title]{Comparison of uncertainty propagation techniques in small-body environment}

\author*[1]{\fnm{Niccolò} \sur{Michelotti}}\email{niccolo.michelotti@polimi.it}

\author[1]{\fnm{Antonio} \sur{Rizza}}\email{antonio.rizza@polimi.it}

\author[1]{\fnm{Carmine} \sur{Giordano}}\email{carmine.giordano@polimi.it}

\author[1]{\fnm{Francesco} \sur{Topputo}}\email{francesco.topputo@polimi.it}

\affil[1]{\orgdiv{Department of Aerospace Science and Technology}, \orgname{Politecnico di
Milano}, \orgaddress{\street{via La Masa, 34}, \city{Milan}, \postcode{20156}, \country{Italy}}}

%%==================================%%
%% Abstract %%
%%==================================%%
\abstract{
Close-proximity exploration of small celestial bodies is crucial for the comprehensive and accurate characterization of their properties. However, the complex and uncertain dynamical environment around them contributes to a rapid dispersion of uncertainty and the emergence of non-Gaussian distributions. Therefore, to ensure safe operations, a precise understanding of uncertainty propagation becomes imperative. In this work, the dynamical environment is analyzed around two asteroids, Apophis, which will perform a close flyby to Earth in 2029, and Eros, which has been already explored by past missions. The performance of different uncertainty propagation methods (Linear Covariance Propagation, Unscented Transformation, and Polynomial Chaos Expansion) are compared in various scenarios of close-proximity operations around the two asteroids. Findings are discussed in terms of propagation accuracy and computational efficiency depending on the dynamical environment. 
By exploring these methodologies, this work contributes to the broader goal of ensuring the safety and effectiveness of spacecraft operations during close-proximity exploration of small celestial bodies. 
}

\keywords{uncertainty propagation, small bodies exploration, non-Gaussian distribution, proximity operations, polynomial chaos expansion}

\maketitle

%%==================================%%
%% Chapters %%
%%==================================%%

\section{Introduction}
The exploration of small celestial bodies, such as asteroids and comets, stands as a rapidly evolving and cutting-edge field in planetary science, driven by growing interest and recent successful missions. Indeed, in recent years, numerous missions have achieved close encounters, rendezvous or landing on small bodies as one of their primary objectives \citep{palomba_past_future_missions}.
The wide range of scientific and engineering motivations that foster the exploration of these celestial bodies can be summed up in three main areas of interest: the study of the Solar System formation, the investigation for future resource utilization, and the demonstration of planetary protection techniques \citep{scheeres2012book}.
Close-proximity operations are crucial to characterize the external morphology of the body and to estimate its mass, gravity field and spin state with greater accuracy than what could be obtained with ground-based observations \citep{scheeres2012book}.

\medskip

These tasks entail peculiar technological challenges due to the highly unknown and uncertain dynamical environment in which the spacecraft operates. Challenges arise from the fact that the gravity field of the central body is weaker and less uniform than the one of a planet and therefore the spacecraft becomes more sensitive to orbital perturbations, such as solar radiation pressure and third-body effects \citep{scheeres_challenges}. The uncertainties in the dynamical environment arise from different sources, which can be classified into three main categories: navigation errors, actuation errors, and dynamic model errors \citep{luo}. 
To overcome these challenges, a paradigm shift towards autonomy is being recently observed \citep{franzese_autonomous_navigation,franzese_argo_autonomous} and accurate modelling of uncertainties, with limited on-board resources, becomes therefore of paramount importance.

\medskip

In literature, several works have analyzed and implemented uncertainty propagation techniques to solve problems related to orbital determination, conjunction assessment, and robust trajectory optimization.
Armellin et al. \cite{armellin_DA} presented a method for nonlinear propagation of uncertainties in orbital mechanics based on differential algebra to study the close encounter of asteroid Apophis with our planet in 2029. Jones et al. \cite{jonesPCE} demonstrated the use of polynomial chaos expansion for the nonlinear, non-Gaussian propagation of spacecraft state uncertainty.
Vittaldev et al. \cite{vittaldev_GMM_PCE} combined polynomial chaos expansion and Gaussian mixture models in a hybrid technique to propagate initial Gaussian uncertainties for space situational awareness applications. Boone et al. \cite{boone_stt} derived the directional state transition tensors method and applied it to orbit state propagation.
Yang et al. \cite{yang_luo_UT} and Boone et al. \cite{booneGMM, boone_stt_robust} implemented robust trajectory optimization algorithms with unscented transformation, Gaussian mixture models, and state transition tensors, respectively.
In addition, comprehensive reviews about uncertainty propagation in orbital mechanics \citep{luo} and on the modelling of uncertainties around small bodies \citep{feng_survey}, have been published.
However, to the authors' knowledge, no actual comparison of the performance of different uncertainty propagation methods in various scenarios of proximity operations around small bodies has been carried out yet. 

\medskip

This study seeks to cover this gap, with a perspective towards autonomous on-board implementation. Specifically, it aims to investigate the potential benefits of employing accurate nonlinear methods for uncertainty propagation considering a trade-off between solution accuracy and computational efficiency.
To achieve this objective, three uncertainty propagation techniques (Linear Covariance Propagation, Unscented Transformation, and Polynomial Chaos Expansion) are implemented and compared across diverse scenarios involving proximity operations around two asteroids, namely Apophis and Eros. These scenarios are chosen to encompass a range of dynamic environments and conditions. Initial uncertainty distributions are propagated along the trajectories using each method, and the results of the propagated distributions and statistical moments are analyzed to establish general criteria for technique selection based on trajectory features and dynamic conditions. 
We believe that the analyses and results presented in this work are instrumental for the relevant community to select the proper uncertainty propagation method depending on the context.

The paper is structured as follows. Section \ref{sec:state_of_art} presents a brief overview of uncertainty propagation in astrodynamics and the methods implemented in this work, Section \ref{sec:dynamical_environment} illustrates the modelling of the dynamical environment around the small bodies, Section \ref{sec:case_studies} defines the case studies and scenarios analyzed, Section \ref{sec:results} shows the results of uncertainty propagation and discusses the performance of each method, and lastly conclusions and suggestions for future works are drawn in Section \ref{sec:conclusion}.

\section{Uncertainty propagation methods}
\label{sec:state_of_art}

An orbital dynamic problem with random state vector $\bm{x}\in \mathbb{R}^n$ can be expressed by the It\^{o} stochastic differential equation \citep{maybeck_ito}.
The exact time evolution of the related probability density function (PDF) $p(\bm{x},t)$ is described by the Fokker-Planck equation (FPE) \citep{fuller_fokker_planck}, which is a partial differential equation that describes the propagation of a PDF and therefore provides a complete statistical description of the system. The PDF $p(\bm{x},t)$ can be characterized by its statistical moments, and in particular, the ones up to the fourth-order describe its shape. Namely, the \textit{mean} is the measure of the central tendency of the distribution, the \textit{covariance} describes the sample dispersion from the mean value, the \textit{skewness} is a measure of the lack of symmetry in the distribution and, and the \textit{kurtosis} is a measure of whether the tails of the PDF are heavier or lighter than the ones of a normal distribution (which is assumed to have null kurtosis) \citep{maybeck_ito}.
For a set of $n$ values $x_i$ of the random variable $x$, the unbiased skewness is computed as \citep{joanes_ske_kurt}:
\begin{equation}
    s = \frac{\sqrt{n\left(n-1\right)}}{n-2} \frac{\frac{1}{n}\sum_{i=1}^{n}\left(x_i-\bar{x}\right)^3}{\left[ \frac{1}{n} \sum_{i=1}^{n} \left( x_i - \bar{x}\right)^2\right]^{3/2}}
    \label{eq:ch2_skewness_unbiased}
\end{equation}
and the unbiased kurtosis centred about zero is computed as \citep{joanes_ske_kurt}:
\begin{equation}
    k = \frac{n-1}{\left(n-2 \right) \left(n-3 \right)} \left( \left( n+1 \right) \frac{\frac{1}{n}\sum_{i=1}^{n} \left(x_i - \bar{x} \right)^4 }{ \left( \frac{1}{n} \sum_{i=1}^{n} \left(x_i - \bar{x} \right)^2 \right)^2} -3(n-1)\right)
    \label{eq:ch2_kurtosis_unbiased}
\end{equation}
where $\bar{x}$ is the mean value of the distribution of random samples $x_i$.
In general, solving the FPE for high-dimension systems is computationally untrackable, leading to the formulation of simplified techniques for approximating the uncertainty propagation in astrodynamics \citep{luo}.
The ones considered in this work are: 1) Linear Covariance Propagation, 2) Unscented Transformation, 3) Polynomial Chaos Expansion. Additionally, a Monte Carlo (MC) simulation is considered as the reference solution, propagating $N = 10^4$ random samples generated with Latin Hypercube sampling \citep{stein_latin_hs}.

\medskip

In the \textbf{Linear Covariance Propagation (LinCov)}, the initial distribution is analytically propagated with a linearized model using the state transition matrix (STM). The mean $\bm{\mu}$ and the covariance $\bm{P}$ of the initial uncertainty distribution are computed as \citep{luo}:
\begin{align}
\begin{aligned}
    & \bm{\mu}(t) = \bm{\Phi}(t,t_0)\bm{\mu}(t_0) \\
    & \bm{P}(t) = \bm{\Phi}(t,t_0)\bm{P}(t_0)\bm{\Phi}^\top(t,t_0)
    \label{eq:ch2-mean_cov_LinCov}
\end{aligned}
\end{align}
where $\bm{\Phi}(t,t_0)$ is the STM from initial epoch $t_0$ to time $t$.
LinCov has been extensively used in astrodynamics to propagate uncertainties in different scenarios, such as quantification of navigation estimation errors \citep{battin_lincov} or analysis of orbital rendezvous problem \citep{geller_lincov}. The main advantages of LinCov are the simplicity and high computational efficiency. However, its accuracy decreases for highly nonlinear systems and long propagation times.

Rizza et al. \cite{rizza_SCP} proposed a conservative linear approximation of the propagated uncertainty distribution. Under the assumption of LinCov, the standard deviation of the position is bounded by $\sigma_r^{max}$ computed as:
\begin{equation}
    \sigma_r^{max} \left(t_i^*\right) = \sqrt{\lambda^{max}\left(\bm{P}_{rr}\left(t_i^*\right)\right)}
    \label{eq:ch-5_LinCov_circle_1}
\end{equation}
where $\lambda_{max}$ is the maximum eigenvalue of the position covariance matrix $\bm{P}_{rr}$ at the specified time instant $t_i^*$.
An analytical upper bound of this approximation can be defined as \citep{rizza_SCP}:
\begin{align}
\begin{aligned}
    \left( \sigma_r^{max} \left(t_i^*\right)\right)^2 \le & \lambda^{max} \left( \right. \bm{\Phi}_{rr}\left(t_i^*,t_{lm,i}\right) \bm{\Phi}_{rr}^\top\left(t_i^*,t_{lm,i}\right) \lambda^{max}\left(\bm{P}_{rr}\left(t_{lm,i}\right) \right)  + \\ 
    & + \bm{\Phi}_{rv}\left(t_i^*,t_{lm,i}\right) \bm{\Phi}_{rv}^\top\left(t_i^*,t_{lm,i}\right) \lambda^{max}\left(\bm{P}_{vv}\left(t_{lm,i}\right)\right) \left. \right)
    \label{eq:ch-5_LinCov_circle_2}
\end{aligned}
\end{align}
And lastly, also the following inequality holds \citep{rizza_SCP}:
\begin{align}
\begin{aligned}
    \left( \sigma_r^{max} \left(t_i^*\right)\right)^2 \le & \lambda^{max}\left( \bm{\Phi}_{rr}\left(t_i^*,t_{lm,i}\right) \bm{\Phi}_{rr}^\top\left(t_i^*,t_{lm,i}\right) \right) \lambda^{max}\left(\bm{P}_{rr}\left(t_{lm,i}\right) \right)  + \\ 
    & + \lambda^{max} \left( \bm{\Phi}_{rv}\left(t_i^*,t_{lm,i}\right) \bm{\Phi}_{rv}^\top\left(t_i^*,t_{lm,i}\right)  \right) \lambda^{max}\left(\bm{P}_{vv}\left(t_{lm,i}\right)\right)
    \label{eq:ch-5_LinCov_circle_3}
\end{aligned}
\end{align}

Eq.\,(\ref{eq:ch-5_LinCov_circle_1}), (\ref{eq:ch-5_LinCov_circle_2}) and (\ref{eq:ch-5_LinCov_circle_3}) define the radius of three concentric circles that overestimate the linearly propagated uncertainty distribution. Note that Eq.\,(\ref{eq:ch-5_LinCov_circle_1}) and (\ref{eq:ch-5_LinCov_circle_2}) give the same result if the initial covariance matrix is diagonal and spherical (i.e., has the same standard deviation along each axis). Furthermore, Eq.\,(\ref{eq:ch-5_LinCov_circle_2}) and (\ref{eq:ch-5_LinCov_circle_3}) have very similar results when the influence of velocity uncertainty is very limited with respect to the position one. 
Note that these conditions are verified in this work, therefore, for the sake of clarity, only the most conservative circle computed with Eq.\,(\ref{eq:ch-5_LinCov_circle_3}) and confidence $3\sigma$ will be included in the results.

\medskip

\textbf{Unscented Transformation (UT)} consists of propagating suitably chosen samples from the initial distribution and through them estimating the probability density function at each future instant. The samples $\bm{x}_i(t_0)\ (i = 1,2,...,P)$, called sigma-points, are deterministically chosen in order to represent the initial mean $\bm{\mu}_0$ and covariance $\bm{P}_0$. Different techniques exist to choose the sigma-points, for instance, the one implemented in this work uses $P = 2n+1$ sigma-points where $n$ is the dimensionality of the stochastic system and computes them as in \cite{merweUT}:
\begin{align}
\begin{aligned}
    & \bm{x}_0(t_0) = \bm{\mu}_0 \hspace{1cm} & & \bm{x}_i(t_0) = \bm{\mu}_0 + \bm{\Delta \xi}_i  & \text{for} \ i = 1,...,2n\\ 
    & \bm{\Delta \xi}_i = \left(\sqrt{n\bm{P}_0}\right)_i  & & \bm{\Delta \xi}_{i+n} = -\left(\sqrt{n\bm{P}_0}\right)_i   &  \text{for} \ i = 1,...,n
\label{eq:ch2-UT-sigma_points}
\end{aligned}
\end{align}
where $\left(\sqrt{n\bm{P}_0}\right)_i$ indicates the $i$-th column of the matrix square root.
Once they have been propagated through the complete nonlinear dynamics, the resulting mean and covariance can be computed at each time instant as \citep{merweUT}:
\begin{align}
\begin{aligned}
    & \bm{\mu}(t) = \sum_{i=0}^{2n} \omega_i^{(\mu)}\bm{x}_i(t)\\
    & \bm{P}(t) = \sum_{i=0}^{2n} \omega_i^{(P)}\left[\bm{x}_i(t) - \bm{\mu}(t)\right]\left[ \bm{x}_i(t)-\bm{\mu}(t)\right]^\top
        \label{eq:ch2-mean_cov_UT}
\end{aligned}
\end{align}
where $\omega_i$ are the weights attributed at each sigma point and are computed as \citep{merweUT}:
\begin{align}
\begin{aligned}
    & \omega_0^{(\mu)} = 0,  \
    & \omega_i^{(\mu)} = \frac{1}{2n}  & \ \text{for} \ i = 1,...,2n \\ 
    & \omega_0^{(P)} =  2,  \
    & \omega_i^{(P)} = \frac{1}{2n}  & \ \text{for} \ i = 1,...,2n  
\label{eq:ch2-UT-sigma_points_weight}
\end{aligned}
\end{align}
The main advantage of the UT is that, by propagating only $2n+1$ samples, it is highly efficient and capable of providing a second-order approximation of the first two moments (mean and covariance). However, it may not be suited for applications in which the PDF becomes non-Gaussian over time.

\medskip

\textbf{Polynomial Chaos Expansion (PCE)} is a sample-based nonlinear method that is capable of providing an approximation of the higher order moments and of the entire PDF at future time instants \citep{luo}. The idea is to represent the approximate solution $ \bm{\hat{x}}(t)$ with a truncated series of orthogonal polynomials $\bm{\Psi}_k(\bm{x}_0)$:
\begin{equation}
    \bm{\hat{x}}(t) = \sum_{k=0}^P \bm{c}_k(t)\bm{\Psi}_k(\bm{x}_0)
    \label{eq:ch2-PCE_series}
\end{equation}
where $\bm{c}_k(t)$ are the unknown coefficients of the expansion.
In the case where $\bm{x}_0$ represent standard Gaussian or uniform random variables, the orthonormal basis $\bm{\Psi}_k(\bm{x}_0)$ is the tensor products of Hermite or Legendre polynomials, respectively.
The number of terms in the series is defined based on the dimensionality $n$ of the random variables $\bm{x}_0$ and on the maximum order $p$ of the polynomial basis:
\begin{equation}
    P = \frac{(p+n)!}{p! \ \! n!}
    \label{eq:ch2-PCE_number_terms}
\end{equation}
The approximation accuracy increases with increasing order of the polynomial basis, until convergence is reached \citep{jonesPCE}. However, Eq.\,(\ref{eq:ch2-PCE_number_terms}) shows that a higher number $n$ of input variables causes the number of series terms $P$ to explode rapidly, consequently PCE suffers from the curse of dimensionality issue. In this work, a six-dimensional $n = 6$ input random variable and a fourth-order polynomial basis $p = 4$ have been considered, leading to $P = 210$ expansion terms.
The generation of the expansion of Eq.\,(\ref{eq:ch2-PCE_series}) requires the computation of the coefficients $\bm{c}_k$. In this work, a \textit{non-intrusive} method has been implemented, using the least-squares regression technique \citep{jonesPCE}.
It is based on propagating a batch of $L$ samples randomly drawn from the initial stochastic state $\bm{x}_0$ and then computing the coefficients $\bm{c}_k(t)$ such that the sum of squares of differences between the approximate solution $ \bm{\hat{x}}_i(t)$ and the solution $\bm{x}_i(t)$ computed with the propagated samples is minimized at each instant:
\begin{equation}
    \bm{c}_k(t) \approx \arg \min_{\widetilde{\bm{c}}_k(t)} \frac{1}{L} \sum_{i=1}^L \left( \bm{x}_i(t) -\sum_{k = 0}^P \widetilde{\bm{c}}_k(t)\bm{\Psi}_k({\bm{x}_0}_i) \right)^2
    \label{eq:ch-2_PCE_least_square}
\end{equation}
which leads to the solution of the standard least-squares regression problem:
\begin{equation}
    \left( \bm{H}^\top \bm{H} \right) \bm{\hat{c}}(t) = \bm{H}^\top \bm{x}(t)
    \label{eq:ch-2_PCE_least_square_solution}
\end{equation}
where each column of the measurement matrix $\bm{H} \in \mathbb{R}^{L\times P}$ contains samples of the $j$-th element of the PCE basis, namely:
\begin{equation}
    \bm{H}_{ij} = \bm{\Psi}_j \left( {\bm{x_0}}_i \right)
    \label{eq:ch-2_PCE_measurement_matrix}
\end{equation}
The optimal number $L$ of samples to be propagated depends on the desired accuracy and previous works assessed this value to be in the interval $L \in [P,2P]$ \citep{jonesPCE}. In this work, $L = 2P$ has been selected. The experimental design of the selection of random samples ${\bm{x}_0}_i$ has been performed with Latin hypercube sampling \citep{stein_latin_hs}.
Once the PCE coefficients $\bm{\hat{c}}_k$ have been computed, the mean and covariance of the propagated uncertain variables can be directly obtained from them as \citep{jonesPCE}:
\begin{align}
\begin{aligned}
    & \bm{\mu}(t) = \bm{\hat{c}}_0(t) \\
    & \bm{P}(t) = \sum_{k = 1}^P \bm{\hat{c}}_k(t) \bm{\hat{c}}_k^\top(t)
    \label{eq:ch2-mean_cov_PCE}
\end{aligned}
\end{align}
In addition, higher order moments (i.e., skewness and kurtosis) and the entire PDF can be estimated by Monte Carlo-like sampling of the truncated expansion in Eq.\,(\ref{eq:ch2-PCE_series}). Note that this sampling is fast since it does not require the integration of the dynamics over time. In this work, $10^4$ samples are generated from the expansion at each time instant.
Compared to Monte Carlo simulation, the PCE converges up to an exponential behaviour with respect to the order of the polynomial basis even in the case of non-Gaussian uncertainty \citep{luo}. 

\section{Spacecraft dynamics}
\label{sec:dynamical_environment}

The relative dynamics of the spacecraft around the small body is described in a quasi-inertial reference frame centred in the asteroid center of mass and aligned with the ecliptic plane, with inertial fixed axes. $X$ and $Y$ axes lie in the ecliptic plane whereas $Z$ is orthogonal to that. 
The equation of motion for the spacecraft can be defined as:
\begin{equation}
    \bm{\Ddot{r}} = \bm{a}_{cb} + \bm{a}_{3^{rd}} + \bm{a}_{SRP}
    \label{eq:ch2-EoM_dynamics}
\end{equation}

The first contribution comes from the gravitational acceleration of the central body $\bm{a}_{cb}$, modelled using the spherical harmonics representation of the gravitational potential $U_{cb}$ \citep{vallado_book}:
\begin{align}
    & \bm{a}_{cb} = \nabla U_{cb} \label{eq:ch-2_central_body_acceleration}\\
    & U_{cb} = \frac{\mu}{r} \sum_{l = 0}^{\infty} \sum_{m = 0}^{l} \left(\frac{R_e}{r}\right)^l \bar{P}_{l,m} ( \sin \ \phi)\left[ \bar{C}_{l,m} \ \cos \ m\lambda + \bar{S}_{l,m} \ \sin \ m\lambda \right]
    \label{eq:ch-2_spherical_harmonic_complete_normalized}
\end{align}
where $\mu$ is the gravitational constant of the central body, $R_e$ the reference equatorial radius, and $r$, $\phi$, $\lambda$ are the satellite distance, latitude and longitude in the body-fixed reference frame, respectively. The harmonic series is characterized by degree $l$ and order $m$, and by the respective normalized coefficients $\bar{C}_{l,m}$ and $\bar{S}_{l,m}$. Lastly $\bar{P}_{l,m}$ represents the associated Legendre functions. Note that $\bm{a}_{cb}$ needs to be rotated from the asteroid's body-fixed reference frame to the inertial one, considering the ecliptic longitude and latitude of the angular momentum of the asteroid used to define the spherical harmonics model.

The second contribution is the third-body effect caused by the Sun or by planets in proximity of the small body. It is modelled as the differential Keplerian gravity between the barycenter of the small body and the spacecraft position \citep{ferrariTrajectoryOptions}:
\begin{equation}
    \bm{a}_{3^{rd}} = (\mu_{3^{rd}} + \mu_{cb}) \frac{\bm{r}_{cb-3^{rd}}}{||\bm{r}_{cb-3^{rd}}||^3} - \mu_{3^{rd}} \frac{\bm{r}_{sc-3^{rd}}}{||\bm{r}_{sc-3^{rd}}||^3}
    \label{eq:ch2-third_body_acc}
\end{equation}
where, $\mu_{cb}$, $\mu_{3^{rd}}$ are the gravitational constants of the central small body and of the other celestial body (i.e., a planet or the Sun), and $\bm{r}_{cb-3^{rd}}$ and $\bm{r}_{sc-3^{rd}}$ are the relative position of the central small body and the spacecraft with respect to the third celestial body, respectively, in the small body centred inertial reference frame. 

The third contribution is the Solar Radiation Pressure (SRP) computed with the \textit{cannonball model} \citep{scheeres2012book}:
\begin{equation}
    \bm{a}_{SRP} = -\frac{(1+\rho)P_0A_{sc}AU}{cM_{sc}}\frac{\bm{r}_{sc-sun}}{||\bm{r}_{sc-sun}||^3}
    \label{eq:ch2-srp_cannonball}
\end{equation}
where the coefficients $P_0$ is the solar flux at 1 AU and $c$ is the speed of light. The value of the reflectance coefficient $\rho$, the equivalent surface area $A_{sc}$, and the mass $M_{sc}$ are related to a representative spacecraft which resembles a 6U Deep Space CubeSat inspired by Milani CubeSat \footnote{See \url{https://www.heramission.space/hera-mission-milani-cubesat} (last visited on May 16, 2024).}. Their values are reported in Table \ref{table:ch2-astronomical_coefficients} together with the astronomical parameters used to model the dynamics.

Ephemerides of celestial bodies and astronomical parameters are retrieved from the DE425s model of SPICE kernels \footnote{See \url{https://naif.jpl.nasa.gov/pub/naif/generic_kernels/spk/} (last visited on May 16, 2024).}, for major celestial bodies, and from  JPL Horizons System \footnote{See \url{https://ssd.jpl.nasa.gov/horizons/} (last visited on May 16, 2024).}, for the asteroids.

\begin{table}[ht]
    \caption{Astronomical parameters and physical properties of a representative 6U DeepSpace CubeSat}
    \label{table:ch2-astronomical_coefficients}
    \centering 
    \begin{tabular}{l c c c}
    \hline
    \textbf{Parameter} & \textbf{Symbol} & \textbf{Value}  & \textbf{Unit} \\
    \hline  \vspace{-0.25cm} \\
    Sun gravitational constant & $\mu_{S}$ & $1.327124\e{11}$ & $\left[ \text{km}^3 \ \text{s}^{-2} \right]$  \\
    Earth gravitational constant & $\mu_{E}$ & $3.986004\e{5}$ & $\left[ \text{km}^3 \ \text{s}^{-2} \right]$  \\
    Moon gravitational constant & $\mu_{M}$ & $4.9028\e{3}$ & $\left[ \text{km}^3 \ \text{s}^{-2} \right]$  \\
    Apophis gravitational constant & $\mu_{a}$ & $2.862328\e{-9}$ & $\left[ \text{km}^3 \ \text{s}^{-2} \right]$  \\
    Eros gravitational constant & $\mu_{e}$ & $4.460241\e{-4}$ & $\left[ \text{km}^3 \ \text{s}^{-2} \right]$  \\
    Astronomical Unit & $AU$ & $1.495978\e{8}$ & $\left[ \text{km}\right]$  \\
    Speed of light & $c$ & $2.997924\e{5}$ & $\left[ \text{km} \ \text{s}^{-1} \right]$  \\
    Solar flux at 1 AU & $P_0$ & $1367$ & $\left[ \text{W}\ \text{m}^{-2}\right]$ \\
    Reflectance coefficient of the s/c & $\rho$ & $0.3$ & $\left[ -\right]$  \\
    Equivalent surface area of the s/c & $A_{sc}$ & $0.5$  & $\left[\text{m}^2\right]$  \\
    Mass of the s/c & $M_{sc}$ & $12$ & $\left[\text{kg} \right]$ \\
    \hline
    \end{tabular}
\end{table}

\section{Case studies}
\label{sec:case_studies}
The selection of case studies for this work takes into account two different asteroids, Apophis and Eros, which are characterized by complementary dynamical environments. Apophis is a small asteroid around which the dynamics is mainly dominated by the SRP during deep-space phase and by the Earth's gravity during the Earth's fly-by. Eros, instead, is a large asteroid around which the dynamics is dominated by the small body's gravity during all proximity operations.

\medskip

\textbf{(99942) Apophis} is a Near-Earth asteroid discovered in 2004 with a mean radius of 168 m. It has drawn significant attention because of its close encounter with Earth, which will happen on April 13th 2029, when the asteroid will perform a flyby at a distance of approximately 37200 km \citep{scheeres_apophis}. This rare and extraordinary conjunction will be of great interest to study how the Earth gravity will affect the surface and the spin state of the asteroid \citep{valvano_apophis}. In addition, Apophis will reach a minimum distance of approximately 95000 km from the Moon. During this period, the dynamical environment in its proximity will rapidly change.
Due to this peculiar trajectory, two scenarios are selected: 1) proximity during deep-space phase, and 2) proximity during Earth's flyby.

\begin{figure}[H]
    \centering
    \subfloat[during deep-space phase\label{fig:ch-3_apophis_deep_space}]{
        \includegraphics[width=0.5\textwidth]{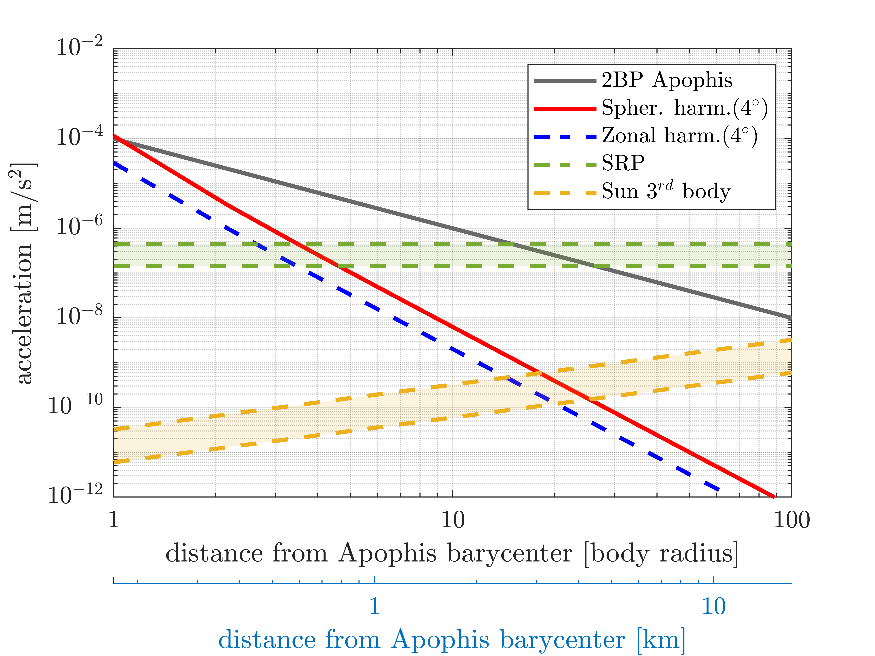}
    }
    \subfloat[during the 3 days of Earth's flyby \label{fig:ch-3_apophis_earth_flyby}]{
        \includegraphics[width=0.5\textwidth]{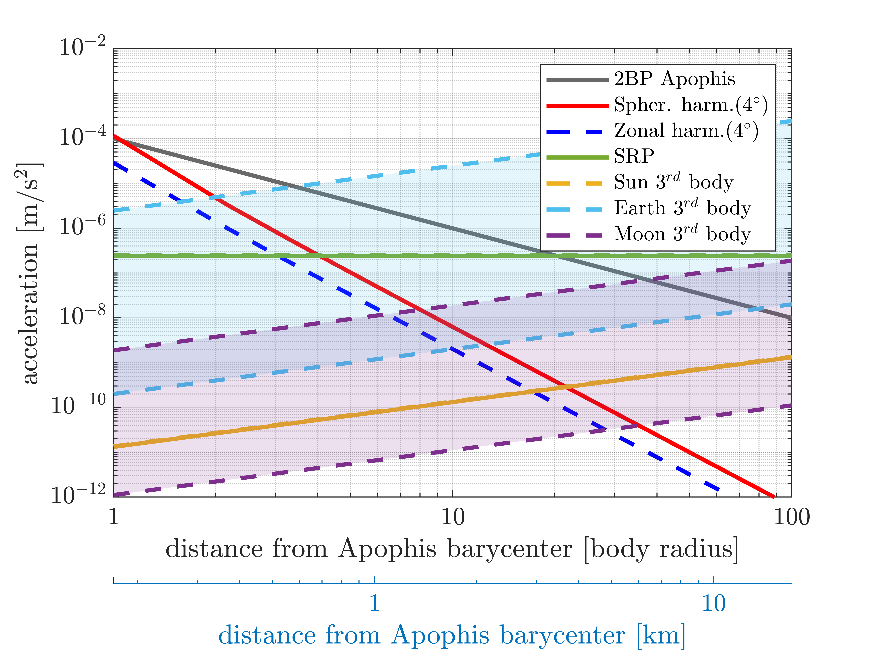}
    }
    
    \caption{Dynamical environment around Apophis. The $x$-axis is shown into two different scales: mean body radius and kilometres. The spherical harmonics model has been implemented up to the fourth order following the one presented in \cite{lang_apophis}}
    \label{fig:ch-3_apophis_dyn_env}
\end{figure}

During the deep-space phase, the main accelerations contributing to the dynamics around Apophis are the asteroid gravity field, the SRP and the Sun third-body effect. 
Figure \ref{fig:ch-3_apophis_dyn_env} shows, for the two scenarios, the magnitude of each perturbation depending on the distance from the target, expressed both in body radius and in kilometres.
Note that, in Figure \ref{fig:ch-3_apophis_deep_space} the acceleration due to SRP and Sun third-body effect are wide bands due to the varying distance of Apophis from the Sun during deep-space phase.
Similarly, Figure \ref{fig:ch-3_apophis_earth_flyby} shows that the third-body effects caused by Earth and Moon vary by several orders of magnitude during the flyby.

\textbf{(433) Eros} is a Near-Earth asteroid with a highly irregular shape and already explored in proximity by the NASA mission NEAR Shoemaker which orbited and landed on the asteroid \citep{yeomans_NEAR}.

\begin{figure}[H]
    \centering
    \includegraphics[width=0.5\textwidth]{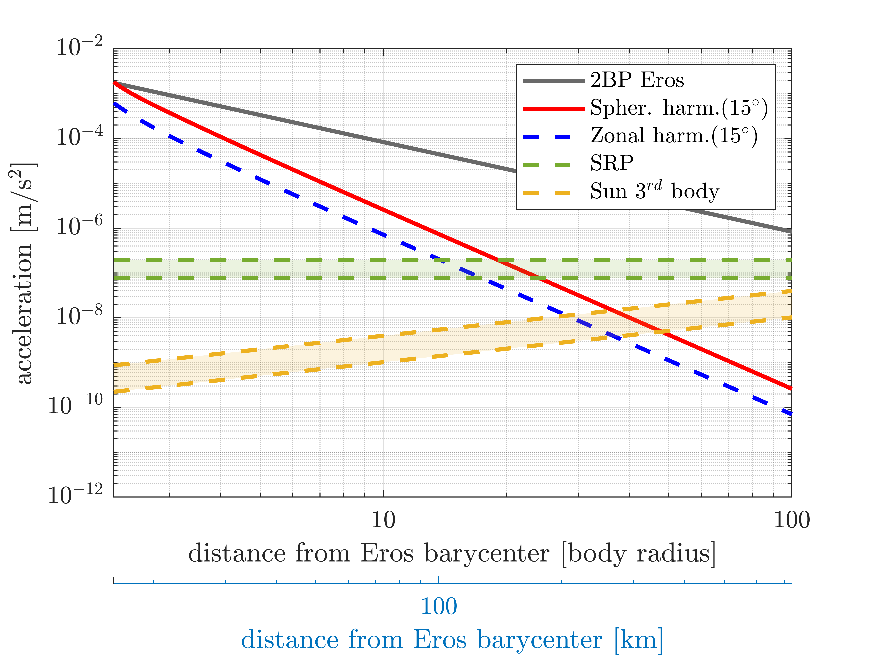}
    \caption{Dynamical environment around Eros along its heliocentric orbit. The $x$-axis is shown into two different scales: mean body radius ($r_m = 7.311$ km) and kilometres. The spherical harmonics model \textit{NEAR15A} is implemented up to the 15th order\protect \footnotemark}
    \label{fig:ch-3_eros_deep_space}
\end{figure}

The dynamical environment around Eros is shown in Figure \ref{fig:ch-3_eros_deep_space}. Differently from the Apophis deep-space scenario, in this case the acceleration of the central body remains dominant even for large distances from the body barycentre, up to more than 1000 km. Moreover, the largest dynamic perturbation is the one deriving from the non-uniform gravity field, described by the spherical harmonics, for distances up to more than 100 km.
\footnotetext{NEAR15A gravity model - See \url{https://sbn.psi.edu/pds/resource/nearbrowse.html} (last visited on May 16, 2024).}

\subsection{Scenarios and initial conditions}

\begin{figure}[H]
    \centering
    \subfloat[\label{fig:ch-5_scenario1_arc_orbits_2km_a}]{
        \includegraphics[width=0.5\textwidth]{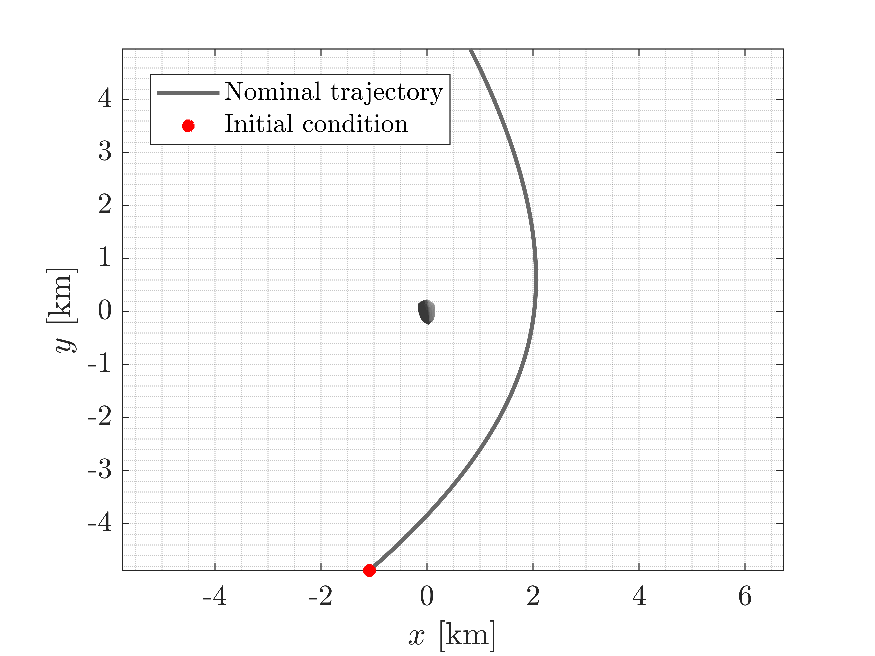}
    }
        \subfloat[\label{fig:ch-5_scenario1_arc_acc_2km}]{
        \includegraphics[width=0.5\textwidth]{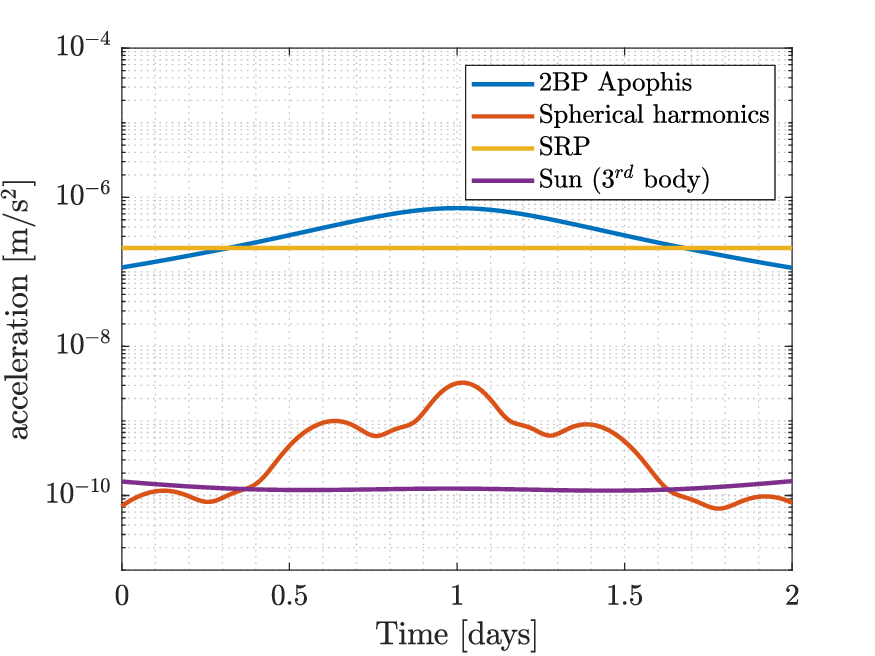}
    }
    \caption{Scenario 1. (a) nominal trajectory, (b) dynamical environment}
    \label{fig:ch-5_scenario1_arc_orbits_2km}
\end{figure}

\begin{figure}[H]
    \centering
    \subfloat[\label{fig:ch-5_scenario1_circ_orbit_a}]{
        \includegraphics[width=0.5\textwidth]{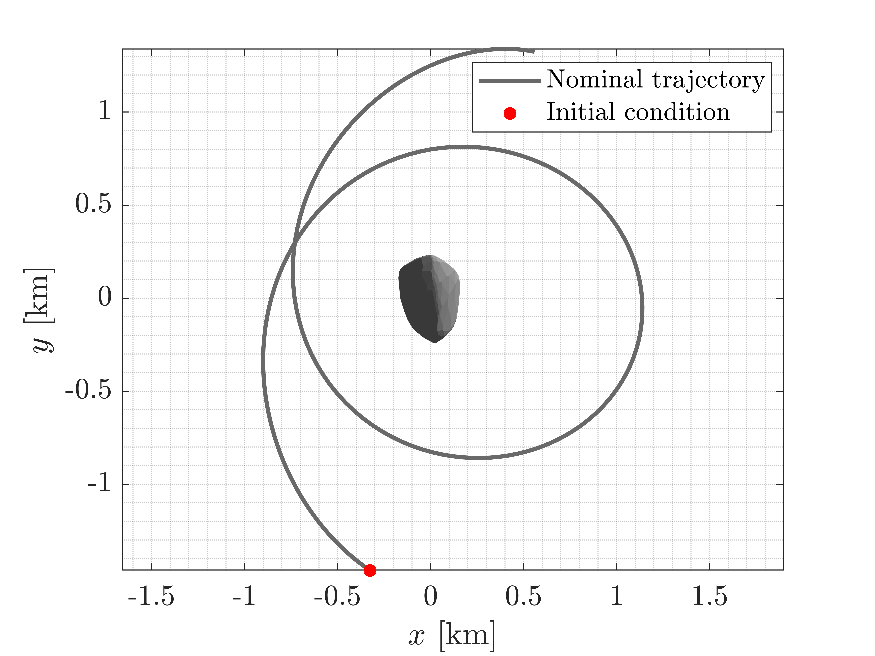}
    }
    \subfloat[\label{fig:ch-5_scenario1_circ_acc}]{
        \includegraphics[width=0.5\textwidth]{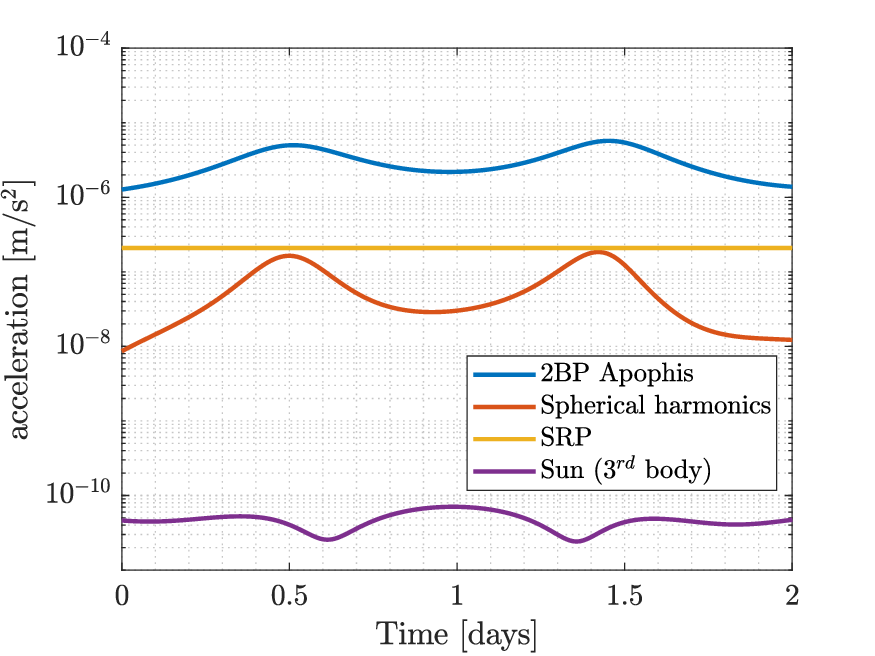}
    }
    \caption{Scenario 2. (a) nominal trajectory, (b) dynamical environment}
    \label{fig:ch-5_scenario1_circ_orbit}
\end{figure}

\begin{figure}[H]
    \centering
    \subfloat[\label{fig:ch-5_scenario2_arc_3km_orbit_a}]{
        \includegraphics[width=0.5\textwidth]{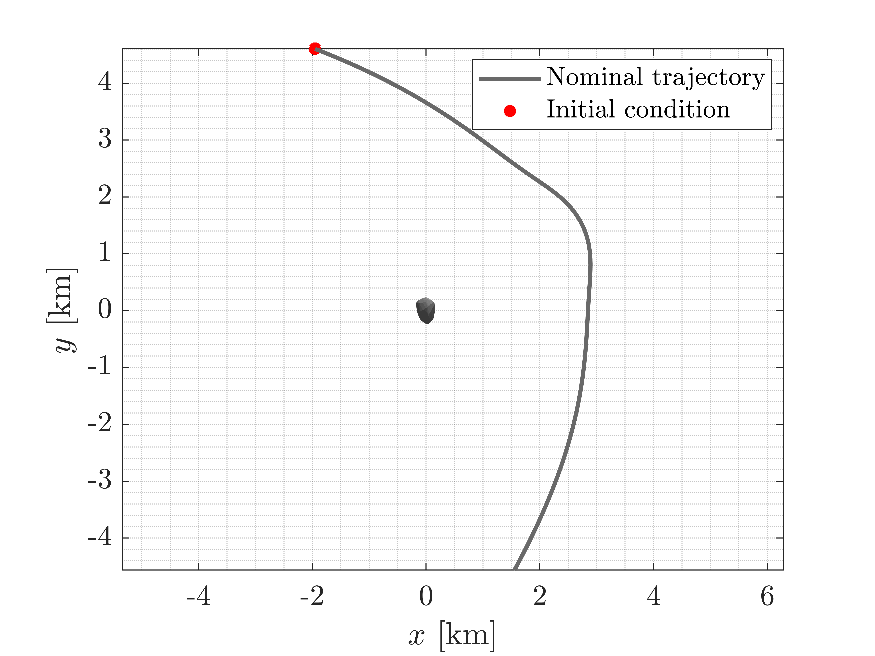}
    }
    \subfloat[\label{fig:ch-5_scenario2_arc_3km_acc}]{
        \includegraphics[width=0.5\textwidth]{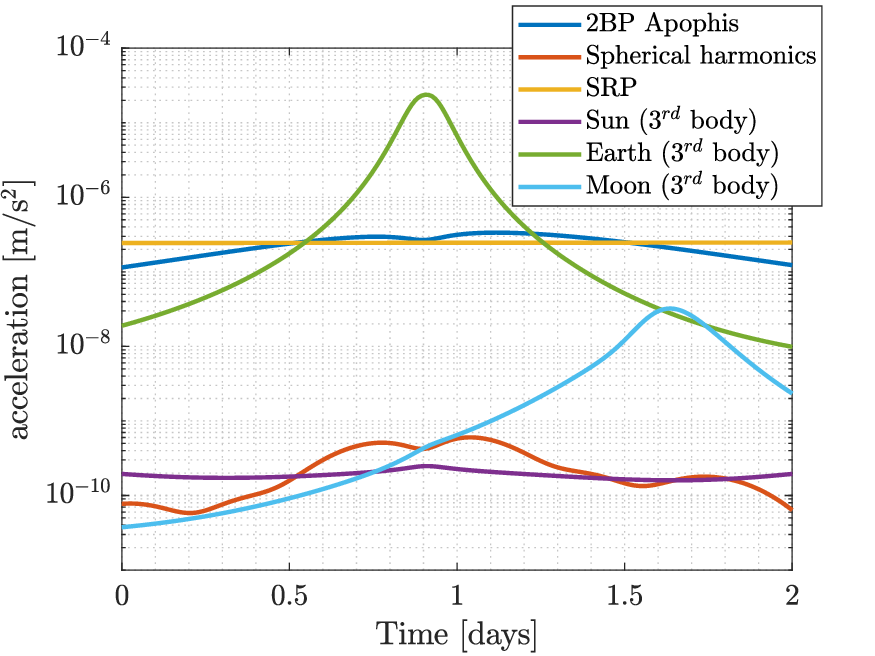}
    }
    \caption{Scenario 3. (a) nominal trajectory, (b) dynamical environment}
    \label{fig:ch-5_scenario2_arc_3km_orbit}
\end{figure}

\begin{figure}[H]
    \centering
    \subfloat[\label{fig:ch-5_scenario2_circ_orbit_a}]{
        \includegraphics[width=0.5\textwidth]{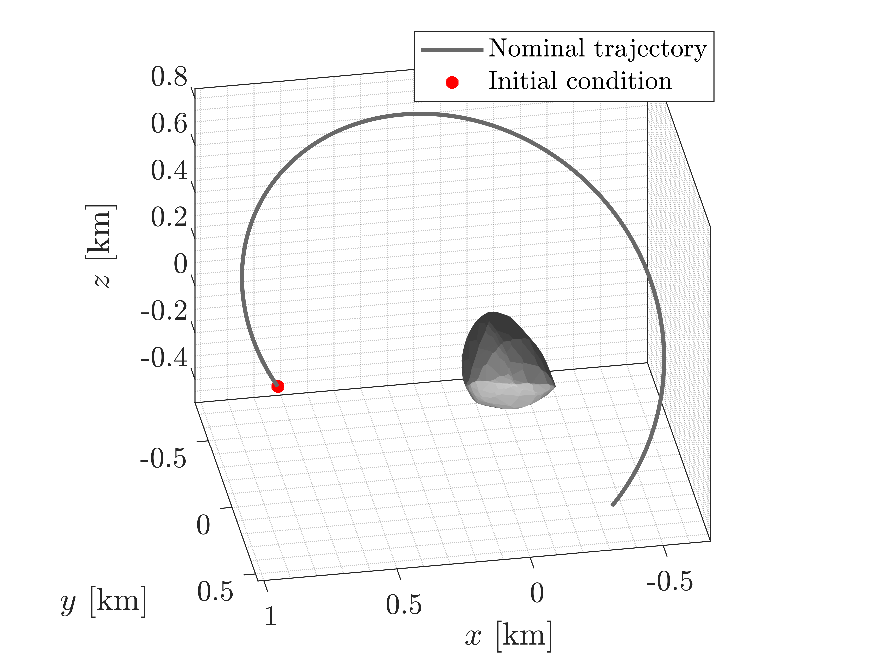}
    }
    \subfloat[\label{fig:ch-5_scenario2_circ_acc}]{
        \includegraphics[width=0.5\textwidth]{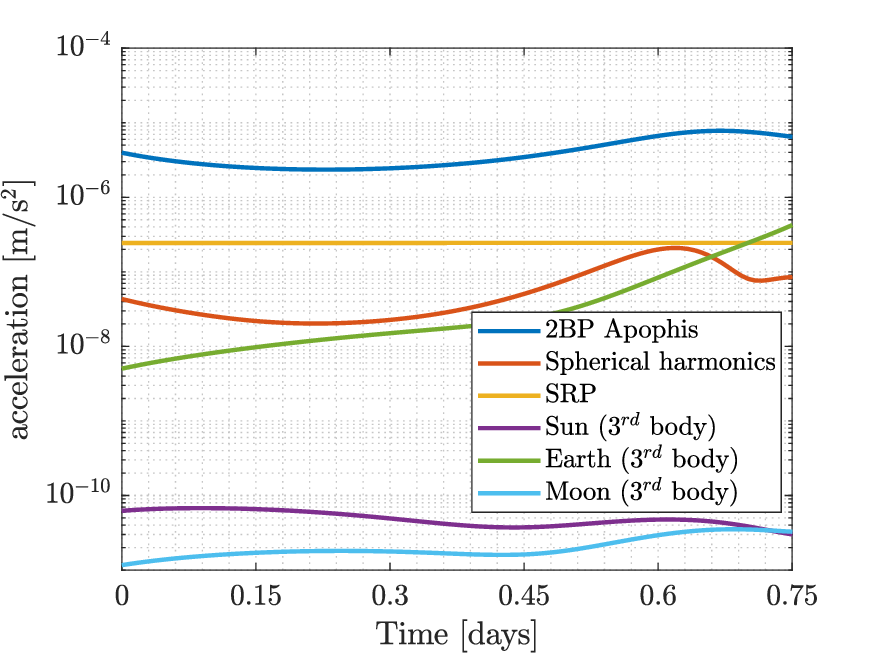}
    }
    \caption{Scenario 4. (a) nominal trajectory, (b) dynamical environment}
    \label{fig:ch-5_scenario2_circ_orbit}
\end{figure}

\begin{figure}[H]
    \centering
    \subfloat[\label{fig:ch-5_scenario3_circ_orbit_a}]{
        \includegraphics[width=0.5\textwidth]{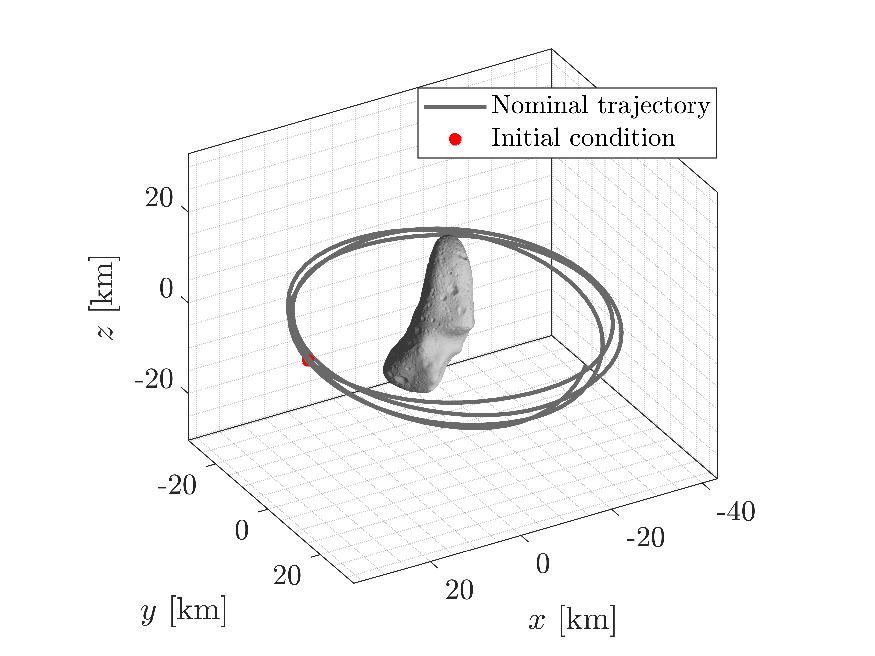}
    }
    \subfloat[\label{fig:ch-5_scenario3_circ_acc}]{
        \includegraphics[width=0.5\textwidth]{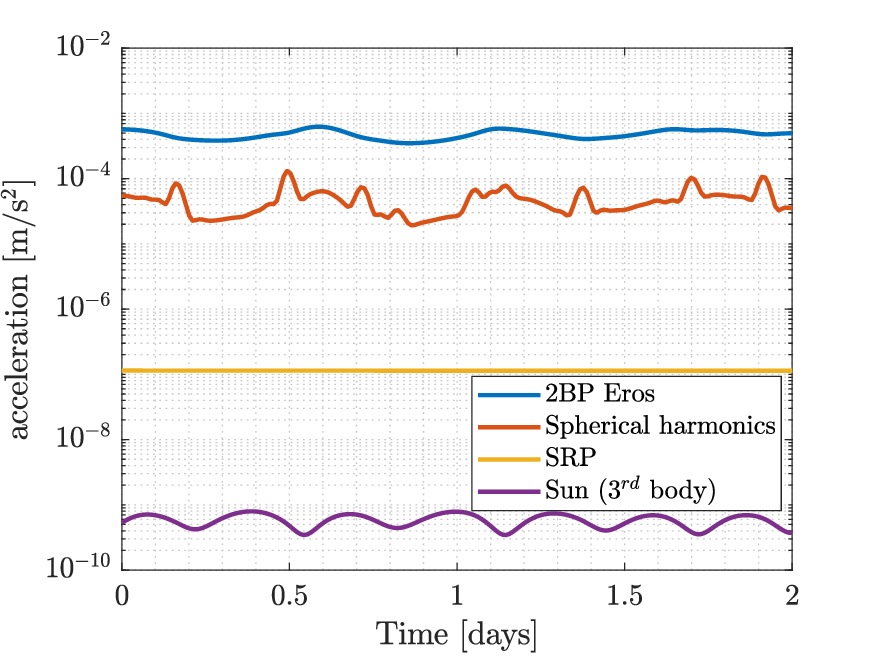}
    }
    \caption{Scenario 5. (a) nominal trajectory, (b) dynamical environment}
    \label{fig:ch-5_scenario3_circ_orbit}
\end{figure}

\begin{figure}[H]
    \centering
    \subfloat[\label{fig:ch-5_scenario3_circ_orbit_large_a}]{
        \includegraphics[width=0.5\textwidth]{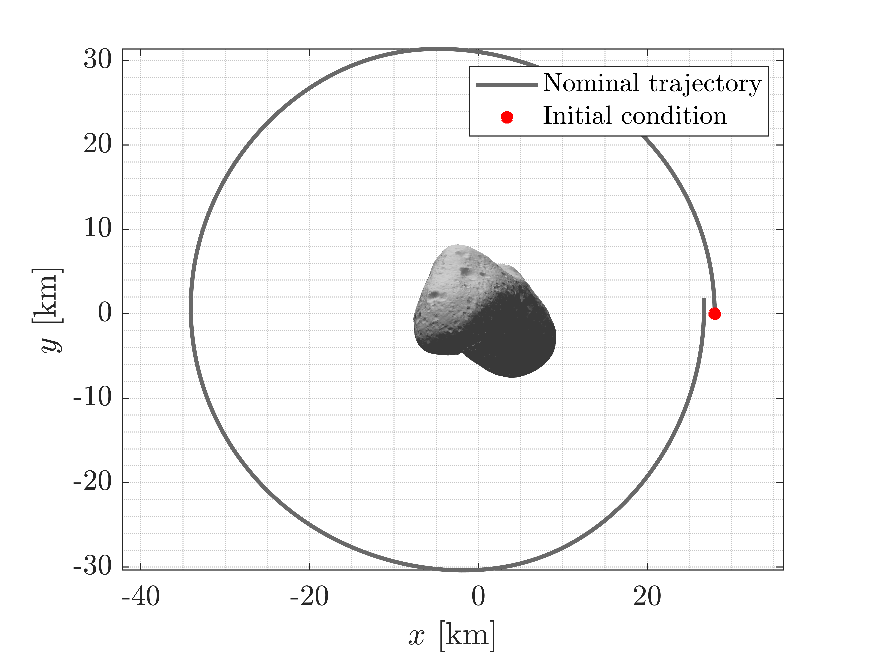}
    }
    \subfloat[\label{fig:ch-5_scenario3_circ_acc_large}]{
        \includegraphics[width=0.5\textwidth]{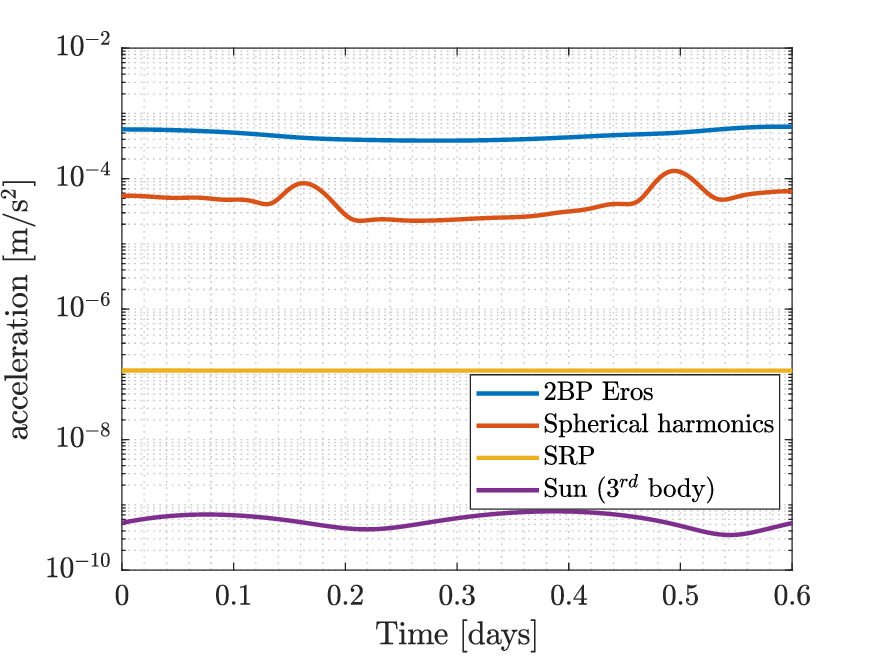}
    }
    \caption{Scenario 6. (a) nominal trajectory, (b) dynamical environment}
    \label{fig:ch-5_scenario3_circ_orbit_large}
\end{figure}

\begin{figure}[H]
    \centering
    \subfloat[\label{fig:ch-5_scenario4_35km_orbit_a}]{
        \includegraphics[width=0.5\textwidth]{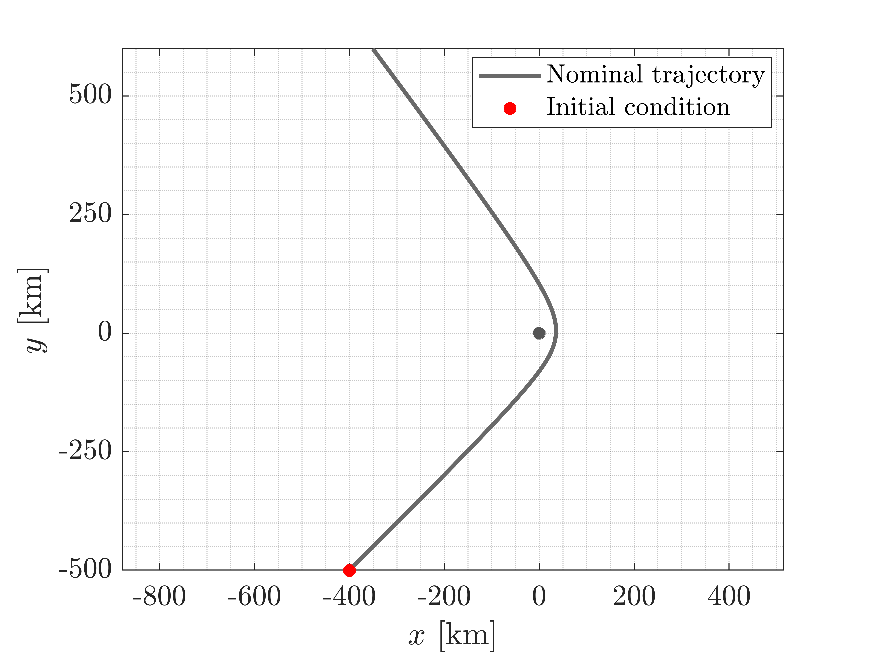}
    }
    \subfloat[\label{fig:ch-5_scenario4_35km_acc}]{
        \includegraphics[width=0.5\textwidth]{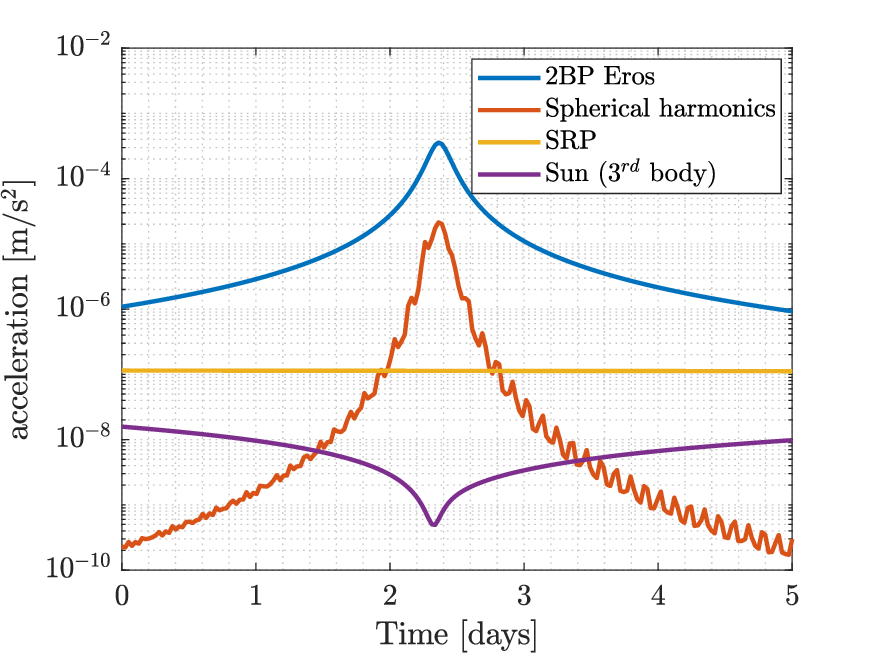}
    }
    \caption{Scenario 7. (a) nominal trajectory, (b) dynamical environment}
    \label{fig:ch-5_scenario4_35km_orbit}
\end{figure}

This work analyzes several scenarios considering different spacecraft trajectories around the two asteroids. The initial state of the spacecraft is modelled as a Gaussian random variable with mean $\bm{x}_0$ and covariance matrix $\bm{P}_0$:
\begin{equation}
    \bm{x}(t_0) \sim \mathcal{N}(\bm{x}_0,\bm{P}_0)
    \label{eq:ch-4_initial_state}
\end{equation}
The covariance matrix $\bm{P}_0$ is assumed spherical with given position and velocity variance $\sigma_{r_0}^2$ and $\sigma_{v_0}^2$. 
Table \ref{table:ch-4_case_studies} reports each scenario analyzed with the related initial conditions. 
In Scenarios from 1 to 5, the initial standard deviations of $\sigma_{r_0} = 10$ m and $\sigma_{v_0} = 0.3$ mm/s along each axis are derived from the expected navigation performance of Hera's Milani CubeSat \citep{bottiglieri_milani,pugliatti_navigation}.
\begin{table}[h!]
    \caption{Scenarios and initial conditions in the quasi-inertial reference frame centred in the asteroid (pericentre radius $r_p$, propagation time $T$)}
    \label{table:ch-4_case_studies}
    \centering
    \begin{tabular}{c l l c c c c c}
   \bottomrule % \hline
         \textbf{\#} & \textbf{Asteroid} & \textbf{Trajectory} & \textbf{$r_p$} & \textbf{$T$} & \textbf{$\sigma_{r_0}$}& \textbf{$\sigma_{v_0}$} & \hspace{0.18cm} \\
                     &                   &            \textbf{description}                      & [km] & [days] & [m]& [mm/s] \\ \hline \vspace{-0.25cm} \\
                     
         1 & Apophis (deep-space) & Hovering arc & 2 & 2 & 10 & 0.3  \\ 
         2 & Apophis (deep-space) & Single revolution &  0.75  & 2 & 10 & 0.3  \\ 
         3 & Apophis (Earth's flyby) & Hovering arc & 3  & 2  & 10  & 0.3 \\ 
         4 & Apophis (Earth's flyby) & Single revolution &  0.55  & 0.75  & 10  & 0.3  \\ 
         5 & Eros & Multiple revolutions &  26 & 2  & 10  & 0.3  \\ 
         6 & Eros & Single revolution &  26  & 0.6  & 30  & 10  \\
         7 & Eros & Flyby &  35  & 5  & 1000  & 10  \\
     \bottomrule
    \end{tabular}
    \smallskip 
        \begin{tabular}{c l l l}
         \textbf{\#} & \textbf{Initial position} & \textbf{Initial velocity} & \textbf{Init. Epoch}\\ & [km] &  [km/s] & [00:00:00 UTC] \\ \hline  \vspace{-0.25cm} \\ 
         1 & $\left[ -1.0850, \ -4.8777, \ 0.1732 \right]^\top$ & $\left[ 4.6808, \ 4.0501, \ -0.015048 \right]^\top 1\mathrm{e}{-5}$ & \texttt{2028 APR 13}\\
         2 & $\left[ -0.3255, \ -1.4633, \ 0.0520 \right]^\top$ & $\left[ -2.8502, \ 1.9168, \ -0.18891 \right]^\top  1\mathrm{e}{-5}$ & \texttt{2028 APR 13}\\
         3 & $\left[ -1.9527, \ 4.6029, \ -0.0042 \right]^\top$ & $\left[ 5.8290, \ -2.2598, \ 1.2331 \right]^\top  1\mathrm{e}{-5}$ & \texttt{2029 APR 13} \\
         4 & $\left[ 0.85, \ 0, \ 0 \right]^\top$ & $\left[ 2.7229, \ -2.1036, \ 4.1277 \right]^\top  1\mathrm{e}{-5}$  & \texttt{2029 APR 13}\\
         5 & $\left[ 28, \ 0, \ 0 \right]^\top$ & $\left[0, \ 0.004, \ 0 \right]^\top$ & \texttt{2028 APR 13}\\
         6 &$\left[ 28, \ 0, \ 0 \right]^\top$ & $\left[0, \ 0.004, \ 0 \right]^\top$  & \texttt{2028 APR 13}\\
         7 & $\left[ -400, \ -500, \ 0 \right]^\top$ & $\left[ 0.002, \ 0.002, \ 0 \right]^\top$ & \texttt{2028 APR 13} \\
     \bottomrule 
    \end{tabular}
\end{table}
Figures from \ref{fig:ch-5_scenario1_arc_orbits_2km} to \ref{fig:ch-5_scenario4_35km_orbit} show for each scenario the nominal trajectory and the contributions to the spacecraft acceleration.
    \vspace{-1.1cm}
    \vspace{-1.1cm}

    \vspace{-1.1cm}
    \vspace{-1.1cm}

Scenarios 1, 2, 3 and 4 are designed to analyze the propagation of uncertainties along trajectories around Apophis which are characterized by different acceleration contributions.
Scenarios 5 and 6 are proximity trajectories around Eros with different initial standard deviations and characterized by significant perturbation due to the distributed gravity field of the asteroid.
Lastly, in Scenario 7, the uncertainty propagation is analyzed during a flyby with relative velocity $\sim 3$ m/s and pericentre altitude of $35$ km. The trajectory has been selected such that the uncertainty distribution and the trajectory itself is affected by the passage nearby Eros, therefore significantly faster and further flybys, which are almost not affected by the asteroid gravity field, have been excluded from this analysis.

\section{Uncertainty propagation results}
\label{sec:results}

An effective way to assess the accuracy of uncertainty propagation is to analyze the propagated covariance matrix of position and velocity at each time instant. In particular, the square root of the trace of the covariance matrix is the metric used in this work to estimate the overall standard deviation of the state, since it does not depend on the adopted reference system.
In order to better visualize the difference in accuracy between the uncertainty propagation methods, the results of LinCov, UT and PCE are compared to the ones of MC simulation by computing the relative error:
\begin{equation}
    \epsilon_{\sigma} = \frac{|\sqrt{Tr(\bm{P}_{UP})} - \sqrt{Tr(\bm{P}_{MC})} |}{\sqrt{Tr(\bm{P}_{MC})}}
    \label{eq:ch-5_relative_error_covariance}
\end{equation}
where $\bm{P}_{UP}$ is the covariance propagated with the three methods, and $\bm{P}_{MC}$ is the covariance propagated with MC simulation. 

In addition, for each scenario, the propagated uncertainty distribution is represented along the trajectory. Results of LinCov and UT are represented with light blue and green covariance ellipses ($3\sigma$ confidence), respectively. The conservative approximation of LinCov, computed with Eq.\,(\ref{eq:ch-5_LinCov_circle_3}), is represented in yellow circles. Lastly, red and black samples are related to PCE and MC. 

\subsection{Scenario 1: hovering arc around Apophis in deep-space}

\begin{figure}[H]
    \centering
    \subfloat[\label{fig:ch-5_scenario1_arc_ellipse}]{
        \includegraphics[width=0.5\textwidth]{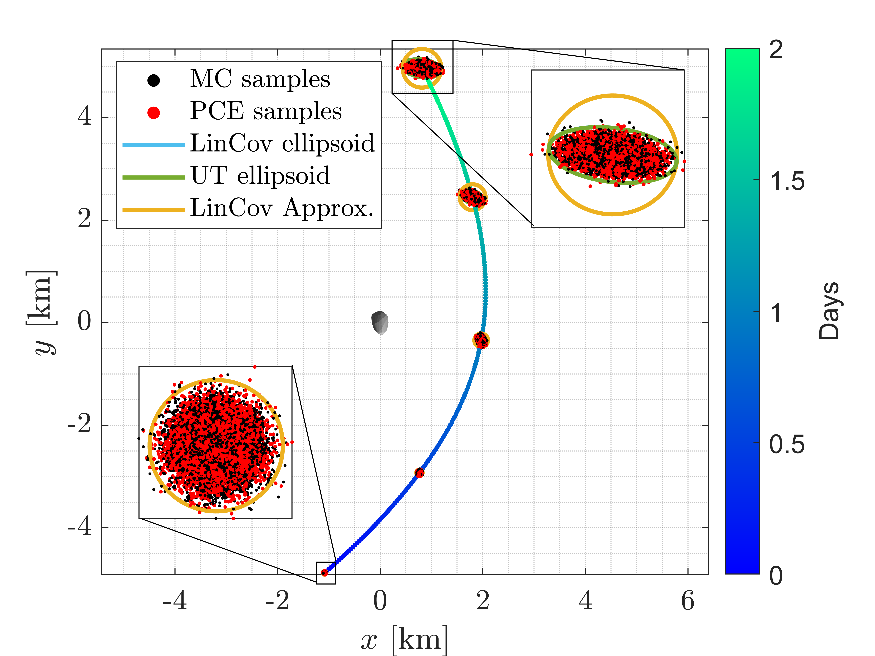}
    }
    \subfloat[\label{fig:ch-5_scenario1_arc_covariance_error}]{
        \includegraphics[width=0.5\textwidth]{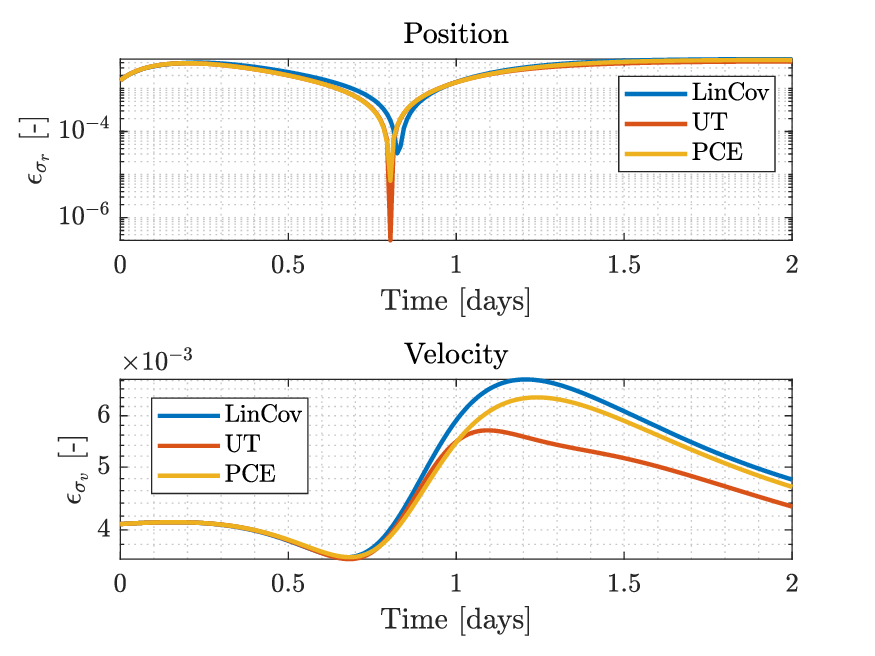}
    }
    \caption{Results of the uncertainty propagation for Scenario 1 in Figure \ref{fig:ch-5_scenario1_arc_orbits_2km}}
    \label{fig:ch-5_scenario1_arc_covariance_error_ellipse}
\end{figure}

For Scenario 1, the relative errors of the estimated standard deviation are reported in Figure \ref{fig:ch-5_scenario1_arc_covariance_error}. The plots show that all the errors are below $10^{-2}$ and considering that the Monte Carlo simulation with $N = 10^4$ samples has an accuracy of approximately $\nicefrac{1}{\sqrt{N}} = 10^{-2}$ \citep{russel_monte_carlo}, all the three methods can be considered to have the same performances in approximating the covariance propagation in this scenario.

The propagated uncertainties related to the position can also be effectively visualized along with the spacecraft trajectory. Figure \ref{fig:ch-5_scenario1_arc_ellipse} shows the time evolution of position uncertainty distribution for the hovering arc considered here. Note that, for the sake of clarity, this plot shows only the in-plane view since the dynamics is almost planar and dispersion along the out-of-plane direction is minimal. The enlargement of the data shows that the shape and dimension of the distribution vary significantly between the initial and the final time instants of the trajectory. However, the distribution remains Gaussian and well approximated by all uncertainty propagation methods. 

\subsection{Scenario 2: single revolution around Apophis in deep-space}

\begin{figure}[H]
    \centering
    \subfloat[\label{fig:ch-5_scenario1_circ_ellipse}]{
        \includegraphics[width=0.5\textwidth]{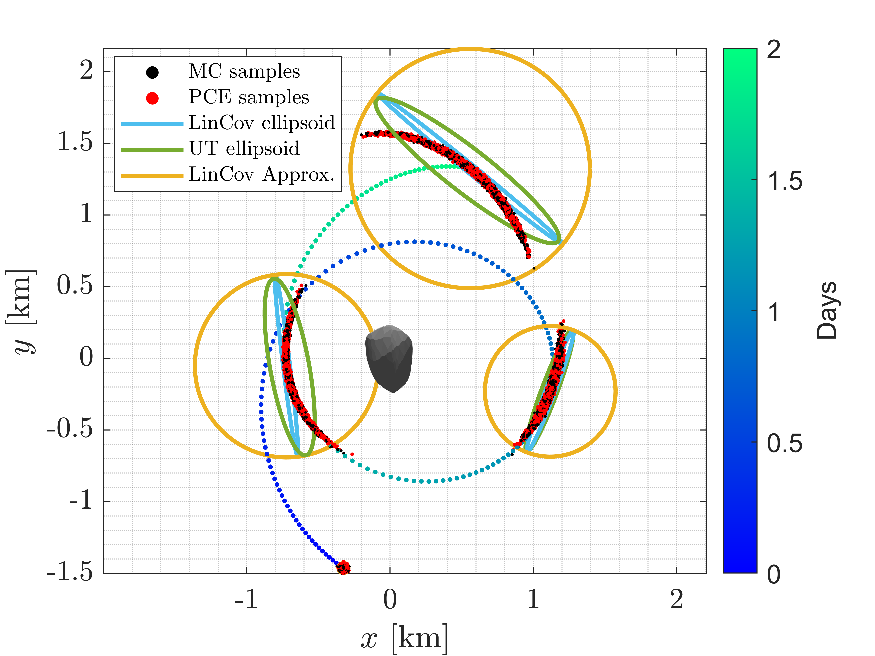}
    }
    \subfloat[\label{fig:ch-5_scenario1_circ_cov_err}]{
        \includegraphics[width=0.5\textwidth]{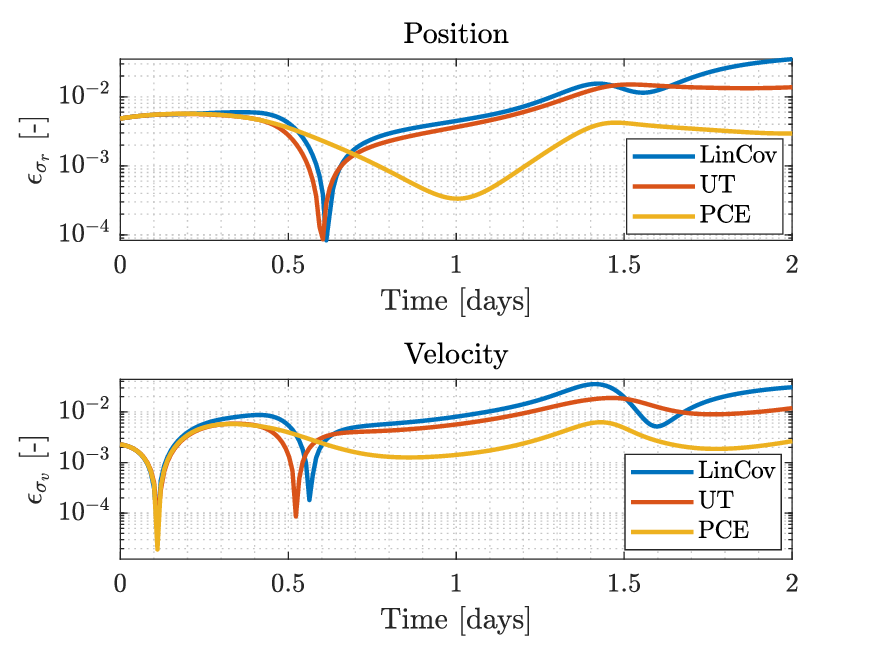}
    }
    \caption{Results of the uncertainty propagation for Scenario 2 in Figure \ref{fig:ch-5_scenario1_circ_orbit}}
    \label{fig:ch-5_scenario1_cov_ellipse}
\end{figure}

The analysis of Scenario 2 in Figure \ref{fig:ch-5_scenario1_cov_ellipse} show that the complex dynamical environment leads to non-Gaussian propagated uncertainty distribution, which assumes a pronounced \textit{banana-shape}. Indeed, Figure \ref{fig:ch-5_scenario1_circ_cov_err} shows that LinCov and UT methods lose accuracy in the covariance propagation, whereas PCE maintains a standard deviation relative error below $10^{-2}$. It is interesting to note that LinCov and UT have similar errors, however the former underestimates the uncertainty dispersion, whereas the latter tends to overestimate it. Indeed, the covariance ellipses of UT include a larger number of MC samples than the ones of LinCov.
The conservative approximation of LinCov largely overestimates the real uncertainty distribution, intersecting also the asteroid surface during close passages.

\begin{figure}[H]
    \centering
    \subfloat[Skewness \label{fig:ch-5_scenario1_circ_skewness}]{
        \includegraphics[width=0.5\textwidth]{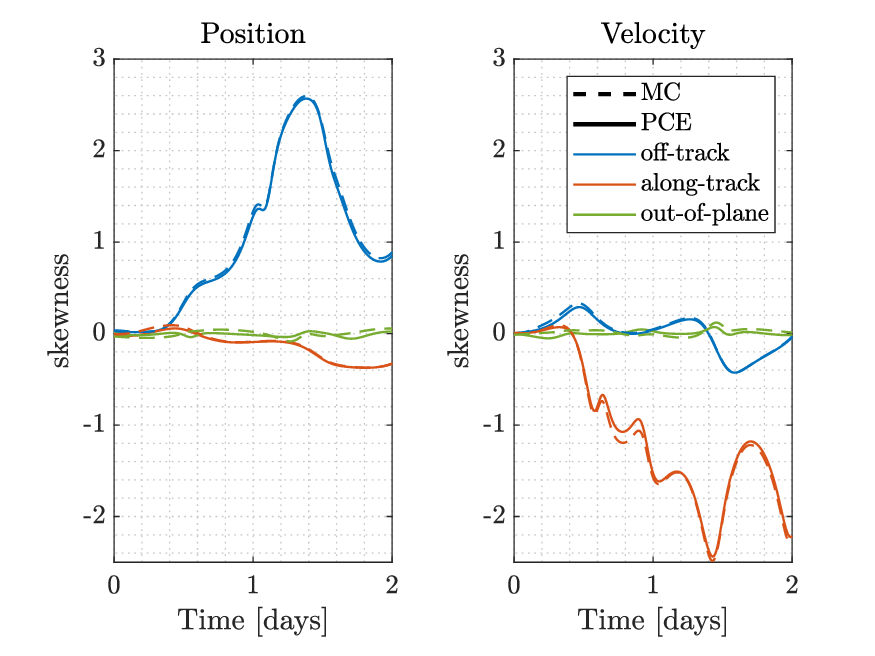}
    }
    \subfloat[Kurtosis \label{fig:ch-5_scenario1_circ_kurtosis}]{
        \includegraphics[width=0.5\textwidth]{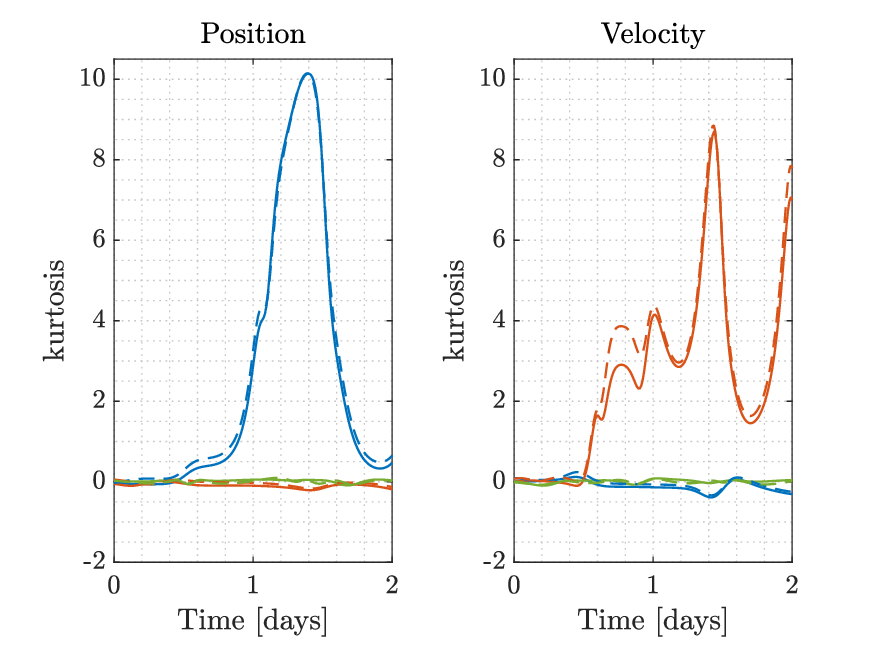}
    }
    \caption{Time evolution of skewness and kurtosis in the RTN reference frame, for Scenario 2 in Figure \ref{fig:ch-5_scenario1_circ_ellipse}}
    \label{fig:ch-5_scenario1_skewness_kurtosis}
\end{figure}

Since the uncertainty distribution becomes non-Gaussian, the analysis of its skewness and kurtosis is of interest to describe the propagated PDF.
As shown in Figure \ref{fig:ch-5_scenario1_skewness_kurtosis}, the higher order moments estimated with PCE (solid lines) are almost the same as the ones derived with MC simulation (dashed lines), showing that this technique is capable of providing an accurate approximation of skewness and kurtosis of the propagated uncertainty distribution. The results are reported in the Radial Tangential Normal (RTN) reference frame, showing that the components with higher values are the normal/off-track component of the position and the tangential/along-track component of the velocity.

\begin{figure}[H]
    \centering
    \subfloat[Histograms for the off-track component of the position \label{fig:ch-5_scenario1_circ_histogram_x}]{
        \includegraphics[width=\textwidth]{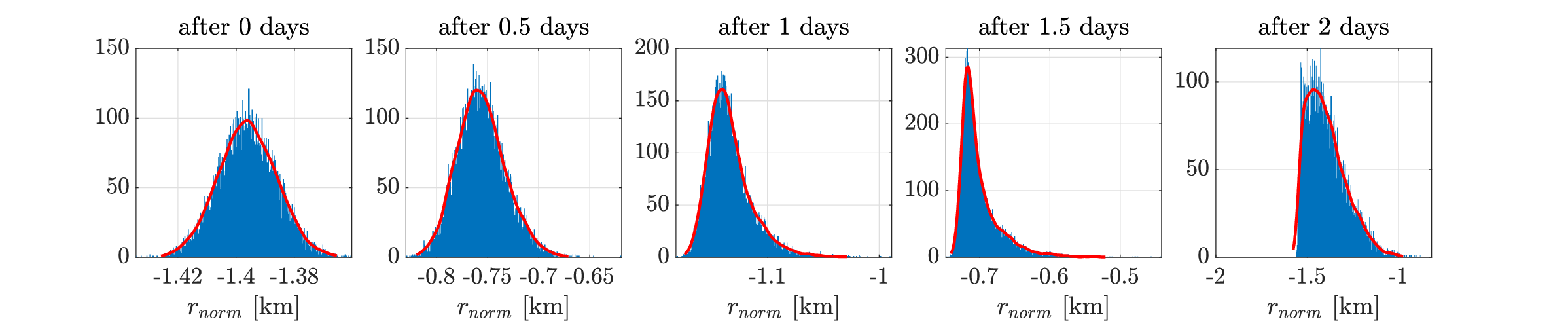}
    }
    \\
    \quad
    \subfloat[Histograms for the along-track component of the velocity \label{fig:ch-5_scenario1_circ_histogram_vx}]{
        \includegraphics[width=\textwidth]{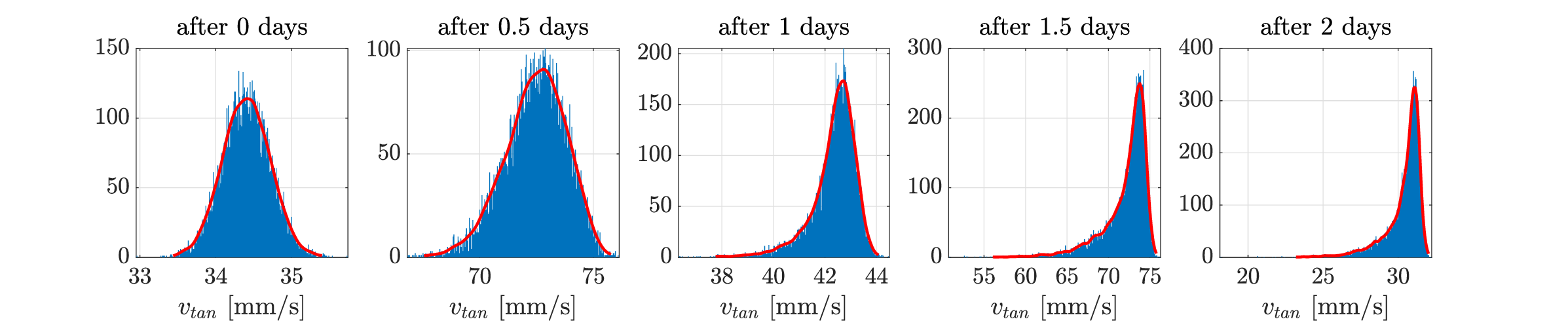}
    }
    \caption{Histograms and kernel density estimation (red solid line \cite{chen_kde}) of the distributions of state components in the RTN reference frame at different time instants, for Scenario 2 in Figure \ref{fig:ch-5_scenario1_circ_ellipse}}
    \label{fig:ch-5_scenario1_histogram}
\end{figure}

Histograms provide a further way to visualize the shape distribution and higher-order statistical moments. In particular, Figure \ref{fig:ch-5_scenario1_histogram} shows histograms at five different time instants of the two most perturbed components of the state in RTN reference frame, that is, the off-track position and the along-track velocity. The plots show clearly how the initial Gaussian distribution evolves into non-Gaussian ones characterized by high values of skewness and kurtosis.

\medskip

\subsection{Scenario 3: hovering arc around Apophis during Earth's flyby}

\begin{figure}[H]
    \centering
    \subfloat[ \label{fig:ch-5_scenario2_arc_3km_ellipse}]{
        \includegraphics[width=0.5\textwidth]{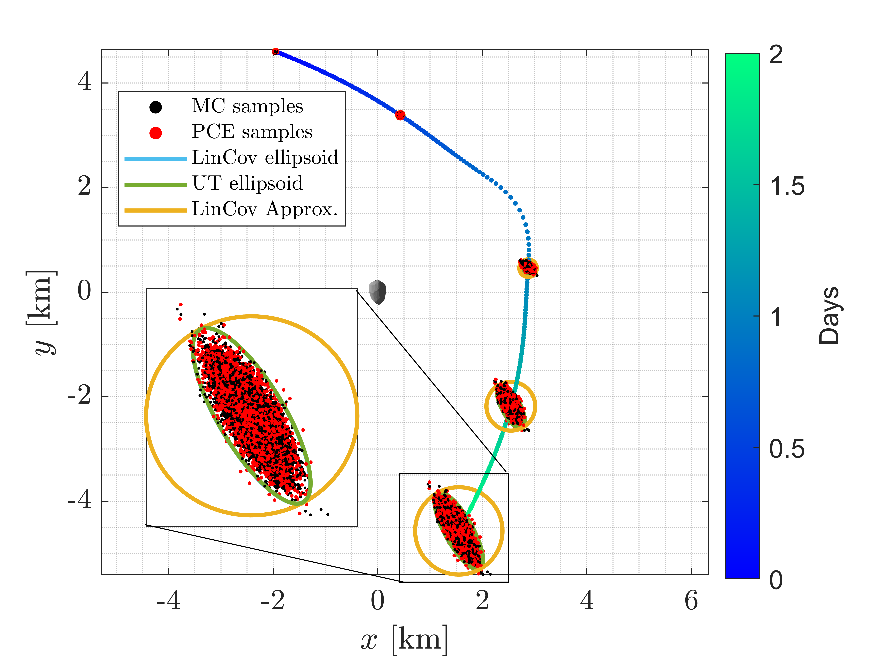}
    }
    \subfloat[\label{fig:ch-5_scenario2_arc_3km_cov_err}]{
        \includegraphics[width=0.5\textwidth]{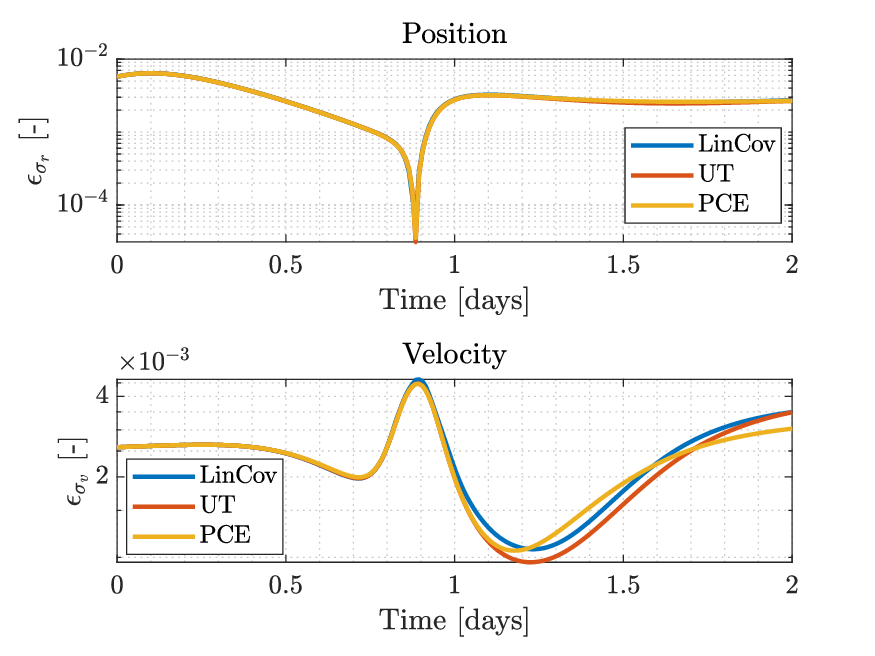}
    }
    
    \caption{Results of uncertainty propagation for Scenario 3 in Figure \ref{fig:ch-5_scenario2_arc_3km_orbit}}
    \label{fig:ch-5_scenario2_arc_3km_cov_ellipse}
\end{figure}

The results of uncertainty propagation for Scenario 3, which considers a hovering arc around Apophis during Earth's flyby, are reported in Figure \ref{fig:ch-5_scenario2_arc_3km_cov_ellipse}. Each method is capable of approximating with good accuracy the propagated uncertainty distribution, which remains essentially Gaussian. Standard deviation relative errors of the three methods are maintained below $10^{-2}$ for the entire propagation time.

\medskip

\subsection{Scenario 4: single revolution around Apophis during Earth's flyby}

\begin{figure}[H]
    \centering
    \subfloat[\label{fig:ch-5_scenario2_circ_ellipse}]{
        \includegraphics[width=0.5\textwidth]{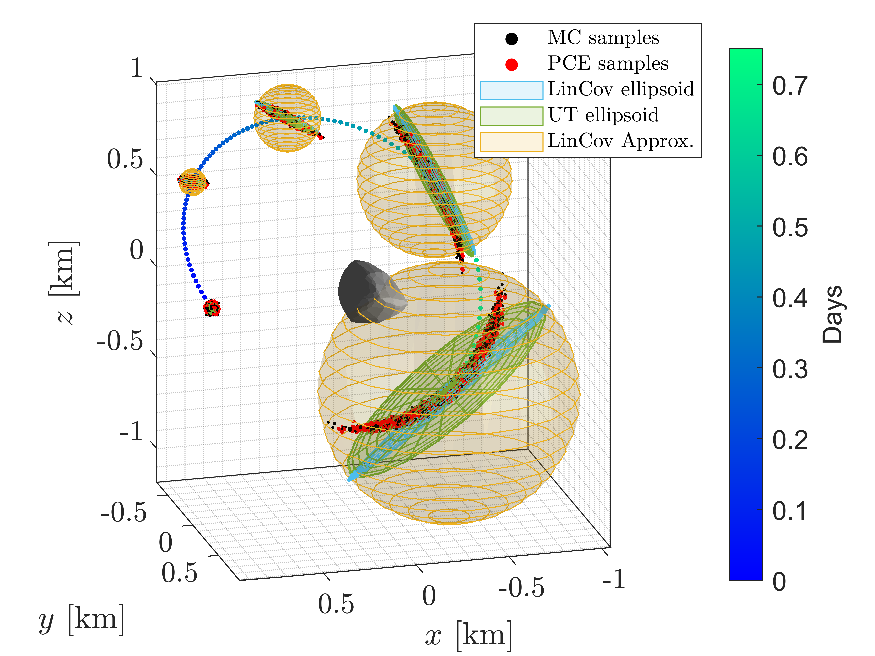}
    }
    \subfloat[ \label{fig:ch-5_scenario2_circ_cov_err}]{
        \includegraphics[width=0.5\textwidth]{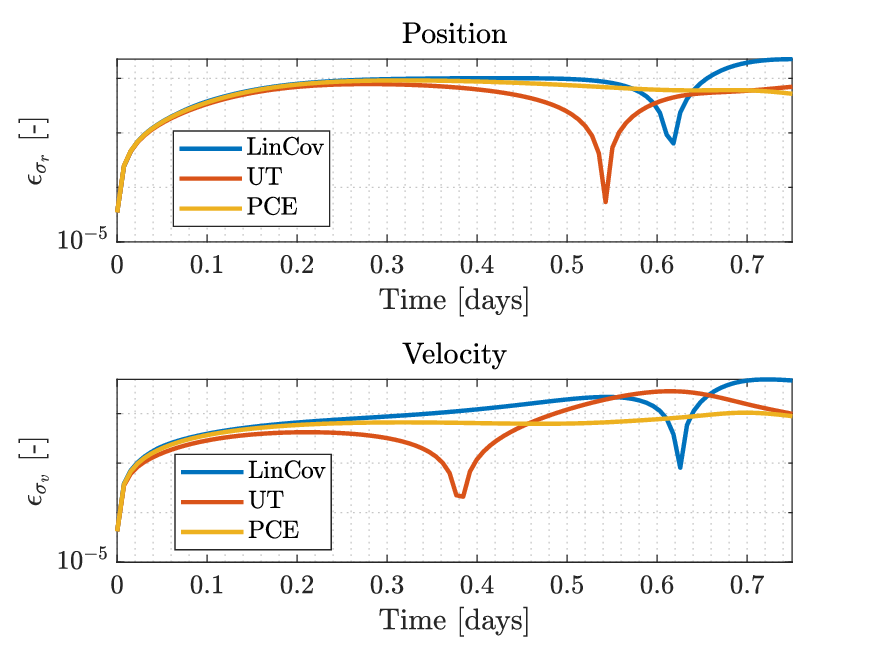}
    }
    \caption{Results of the uncertainty propagation for Scenario 4 in Figure \ref{fig:ch-5_scenario2_circ_orbit}}
    \label{fig:ch-5_scenario2_circ_cov_ellipse}
\end{figure}

Results of Scenario 4 are reported in Figure \ref{fig:ch-5_scenario2_circ_cov_ellipse}, showing that LinCov, UT and, to a lesser extent, PCE show increasing standard deviation relative errors during the trajectory. This is due to the fact that the uncertainty distribution is largely dispersed due to the combined effect of the increasing accelerations of Earth's third body effect and spherical harmonic perturbation, as shown in Figure \ref{fig:ch-5_scenario2_circ_acc}.
In this scenario, PCE is the only method capable to provide an accurate approximation of the propagated uncertainty distribution. 
Note that the conservative approximation of LinCov largely overestimates the uncertainty distribution and intersects the asteroid surface in the final part of the trajectory.

\vspace{-0.1cm}

\subsection{Scenario 5: multiple revolutions around Eros}

\begin{figure}[H]
    \centering
    \subfloat[ \label{fig:ch-5_scenario3_circ_ellipse}]{
        \includegraphics[width=0.5\textwidth]{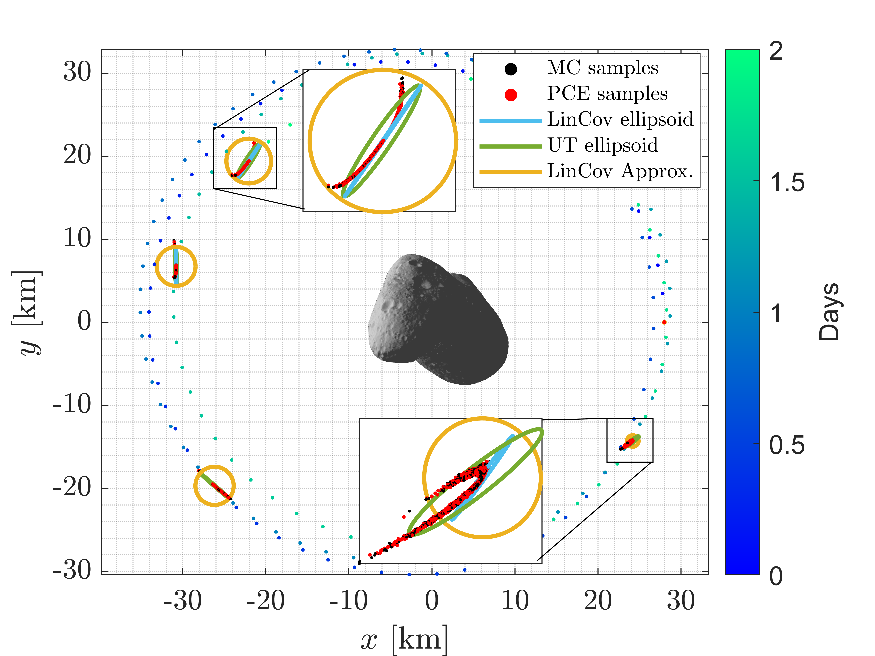}
    }
    \subfloat[\label{fig:ch-5_scenario3_circ_cov_err}]{
        \includegraphics[width=0.5\textwidth]{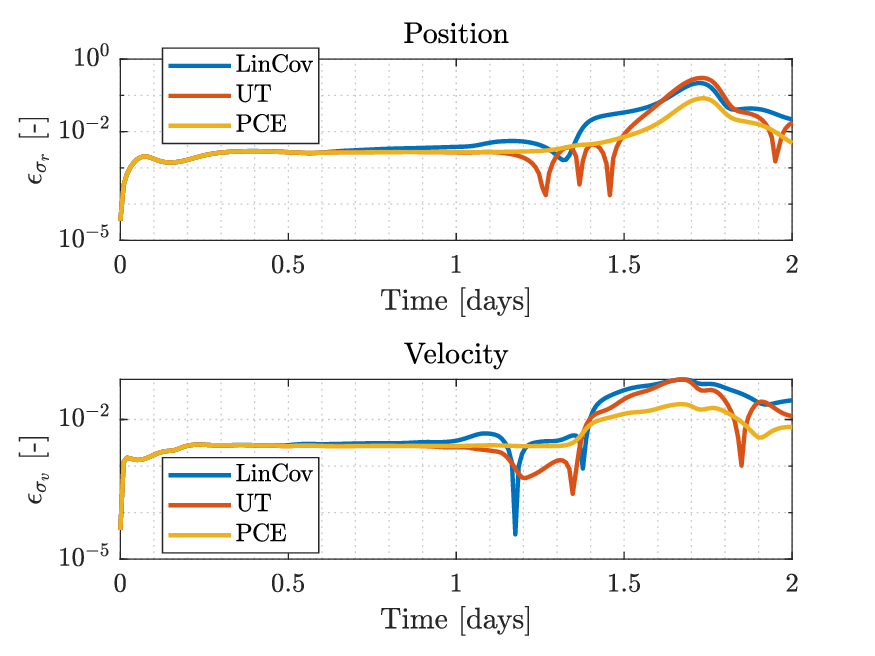}
    }
    \caption{Results of the uncertainty propagation for Scenario 5 in Figure \ref{fig:ch-5_scenario3_circ_orbit}}
    \label{fig:ch-5_scenario3_circ_cov_and_err_ellipse}
\end{figure}

Results of the uncertainty propagation for Scenario 5 are reported in Figure \ref{fig:ch-5_scenario3_circ_cov_and_err_ellipse}. The evolution of standard deviation relative errors shows that all three methods lose accuracy after some close passages to the asteroid surface. However, PCE maintains a better approximation than the other two methods, which reach similar relative errors.
The maximum value of position standard deviation is $\sigma_r = \sqrt{Tr(P_{rr})} \approx 0.7$ km, which is limited with respect to the orbit with radius of approximately 26 km.

Figure \ref{fig:ch-5_scenario3_circ_ellipse} shows enlargements in the position uncertainty distributions at the time instants in which the standard deviation relative error is maximum and at the end of the propagation time. Note that PCE shows an excellent approximation of MC samples, UT tends to overestimate the dispersion of samples, and LinCov underestimates it.  
Lastly, note that in some conditions in which the dynamics is highly nonlinear, even the conservative approximation of LinCov is not able to overestimate the real distribution, because of its complex shape. 

\begin{figure}[H]
    \centering
    \subfloat[Skewness \label{fig:ch-5_scenario3_circ_skewness}]{
        \includegraphics[width=0.5\textwidth]{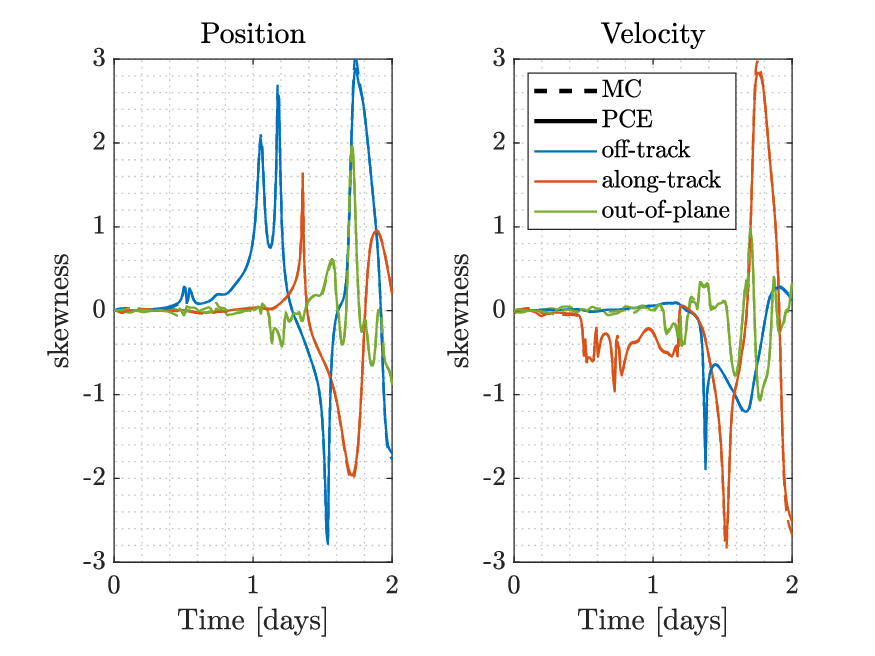}
    }
    \subfloat[Kurtosis \label{fig:ch-5_scenario3_circ_kurtosis}]{
        \includegraphics[width=0.5\textwidth]{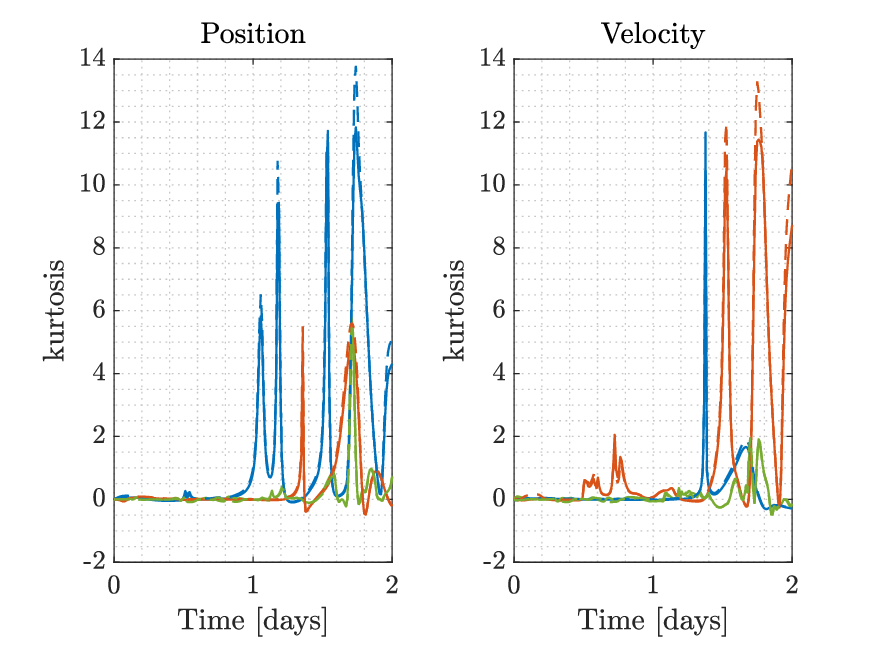}
    }
    \caption{Time evolution of skewness and kurtosis in the RTN reference frame for Scenario 5 in Figure \ref{fig:ch-5_scenario3_circ_ellipse}}
    \label{fig:ch-5_scenario3_skewness_kurtosis}
\end{figure}

Figure \ref{fig:ch-5_scenario3_skewness_kurtosis} shows the skewness and kurtosis of position and velocity in the RTN reference frame. The peaks in higher-order moments correspond to the conditions in which Eros's semi-major axis is perpendicular to the radial direction towards the spacecraft. In this case, the perturbation of the spherical harmonics is maximized since the gravitational field is highly non-uniform and different from the one of a point-mass. This condition largely affects the shape of the propagated uncertainty distribution.

\subsection{Scenario 6: single revolution around Eros}

\begin{figure}[H]
    \centering
    \subfloat[ \label{fig:ch-5_scenario3_larger_covariance_new}]{
        \includegraphics[width=0.5\textwidth]{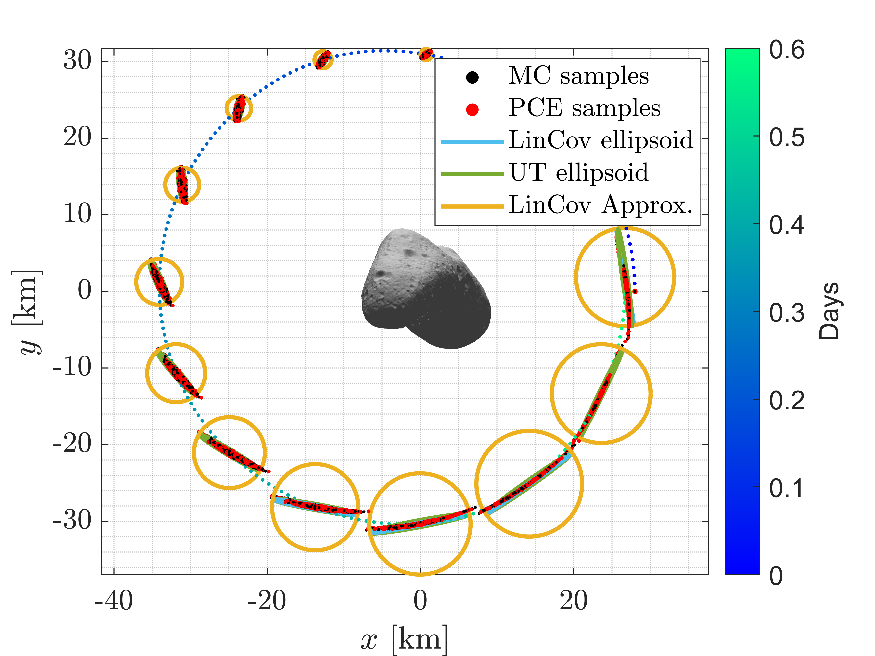}
    }
    \subfloat[\label{fig:ch-5_scenario3_circ_large_cov_err}]{
        \includegraphics[width=0.5\textwidth]{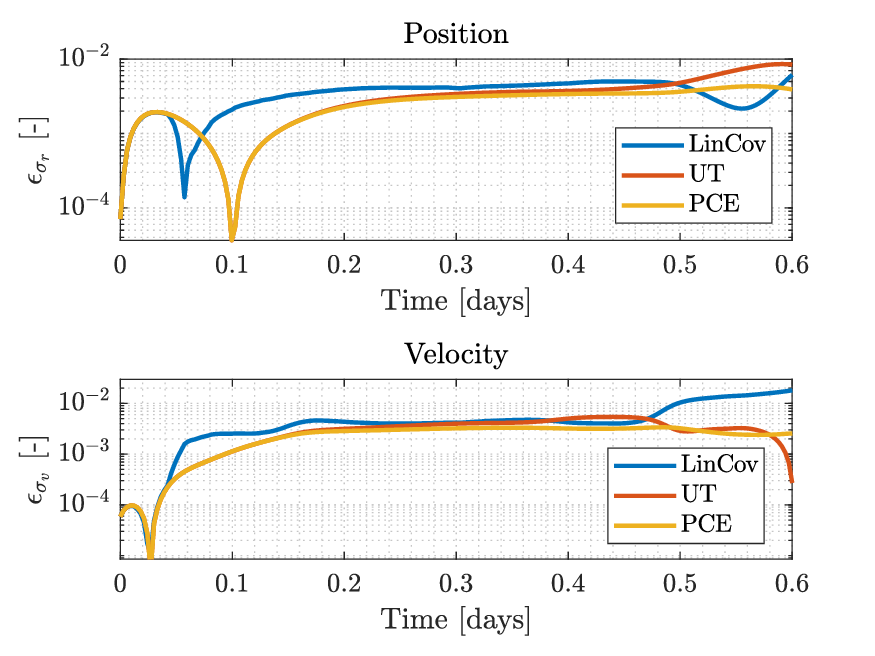}
    }
    \caption{Results of the uncertainty propagation for Scenario 6 in Figure \ref{fig:ch-5_scenario3_circ_orbit_large}}
    \label{fig:ch-5_scenario3_circ_large_cov_and_err}
\end{figure}

Scenario 6 considers larger initial standard deviations, with respect to Scenario 5, and a reduced reference time horizon of a single revolution around the body, which is about 0.6 days. As shown in Figure \ref{fig:ch-5_scenario3_larger_covariance_new}, the conservative approximation of LinCov is always able to overestimate with good accuracy the true non-Gaussian distribution.

\subsection{Scenario 7: low-altitude flyby to Eros}

\begin{figure}[H]
    \centering
    \subfloat[\label{fig:ch-5_scenario4_35km_ellipse}]{
        \includegraphics[width=0.5\textwidth]{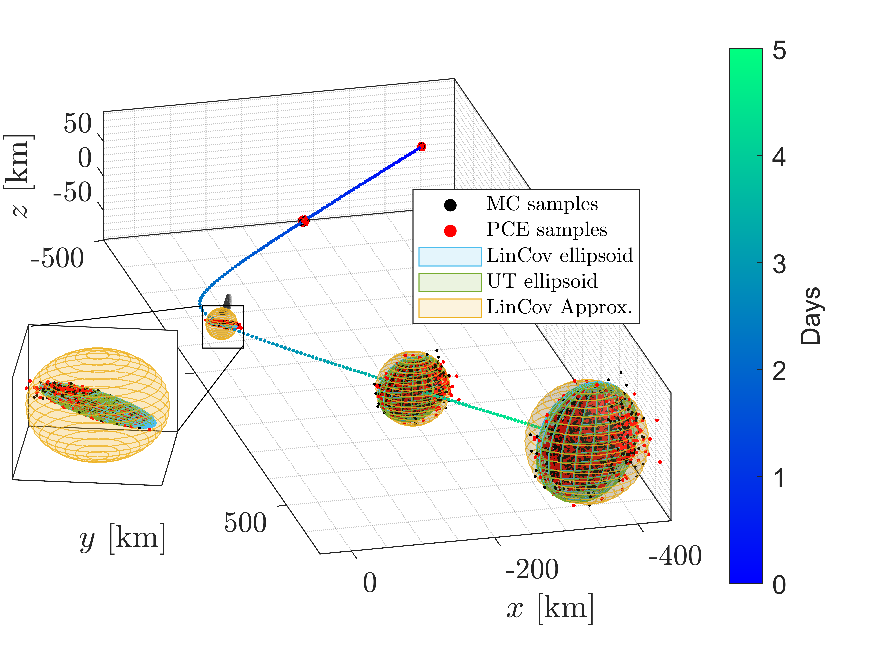}
    }
    \subfloat[ \label{fig:ch-5_scenario4_35km_cov_err}]{
        \includegraphics[width=0.5\textwidth]{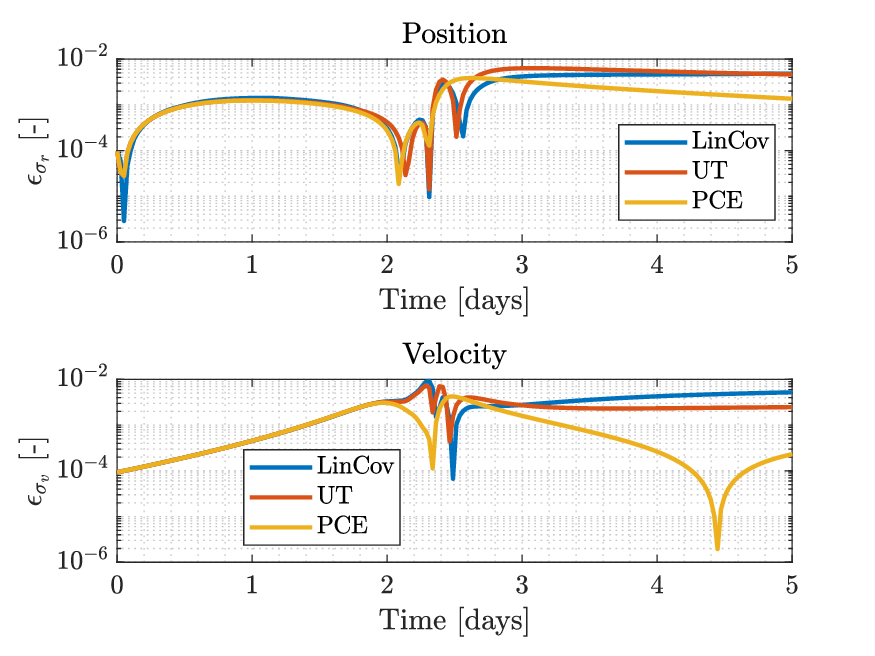}
    }
    \caption{Result of the uncertainty propagation for Scenario 7 in Figure \ref{fig:ch-5_scenario4_35km_orbit}}
    \label{fig:ch-5_scenario4_cov_err}
\end{figure}

The results of uncertainty propagation in Scenario 7 are reported in Figure \ref{fig:ch-5_scenario4_cov_err}.
Despite small differences in the standard deviation relative errors, all the methods approximate with good accuracy the shape of the propagated uncertainty distribution, shown in Figure \ref{fig:ch-5_scenario4_35km_ellipse}. 
Note that differently from the previous scenarios, in this case, the dispersion of the uncertainty distribution is highly significant also in the out-of-plane direction.

\subsection{Discussion on the performance of uncertainty propagation methods}
The uncertainty propagation methods implemented in this work required different computational efforts. Table \ref{table:ch-5_computational_time} shows CPU time required to run the analyzed Scenarios on \texttt{MATLAB-R2023a} (notebook CPU 4Cores/8Threads @2.60GHz and RAM 16GB). Note that the execution time varies significantly depending on the propagation time, the complexity of the dynamics, the accuracy of the integrator, and the efficiency of the code implementation.

\begin{table}[h!]
    \caption{CPU time required by the different uncertainty propagation methods in the scenarios analyzed}
    \label{table:ch-5_computational_time}
    \centering
    \begin{tabular}{l l l}
        \hline
         \textbf{Method} & \textbf{CPU time [s]} & \textbf{Note}  \\
         \hline
         LinCov & $\sim 5-30$ & \makecell[l]{Numerical estimation of the STM \\ with central finite differences}\\ 
         UT & $\sim 30-80$ & $2n+1 = 13$ samples \\
         PCE & $\sim 500-1200$ & $4^\circ$ order PCE with $N = 420$ samples \\
        MC & $\sim 7000-15000$ & $N = 10^4$ samples \\
        \hline
    \end{tabular} 
\end{table}

Some general considerations are reported below to provide guidelines to help in the selection of the most suitable method for each scenario in future works.

\medskip

\textbf{LinCov} is characterized by high computational efficiency and ease of implementation. Depending on the scenario, different performances are shown:
\begin{itemize}
    \item Accurate uncertainty propagation with LinCov has been achieved for simple trajectories such as hovering arcs and flybys, even at low altitudes. 
    \item LinCov results are not accurate for uncertainty propagation in close proximity operations characterized by high nonlinearities and perturbations, such as trajectories which involve one or multiple low-altitude revolutions around the asteroid. Indeed, LinCov performs a linear approximation of the dynamics and it is capable of approximating only Gaussian distributions. In particular, LinCov tends to underestimate the covariance of the uncertainty dispersion in these complex scenarios.
    \item Conservative estimates of the linearly propagated uncertainty distribution, computed with Eq.\,(\ref{eq:ch-5_LinCov_circle_1}), (\ref{eq:ch-5_LinCov_circle_2}) and (\ref{eq:ch-5_LinCov_circle_3}), remain valid for moderately non-Gaussian distributions, such as the \textit{banana-shape}, yet lose accuracy for more complex and highly non-Gaussian distributions.
\end{itemize}

\medskip

\textbf{UT} is characterized by simple implementation and high computational efficiency, as it requires the propagation of only a few samples. Depending on the scenario, UT is characterized by varying performance:
\begin{itemize}
    \item Good accuracy is achieved by UT in the uncertainty propagation for simple trajectories, such as hovering arcs or flybys, even at low altitudes.
    \item The accuracy of UT decreases in case of highly perturbed trajectories, such as multiple revolutions around small bodies, which lead to non-Gaussian uncertainty distribution. Indeed, UT is only able to propagate mean and covariance and therefore only approximates Gaussian distributions. However, the covariance propagated with UT generally overestimates the real uncertainty dispersion, and therefore this method could be implemented in conservative analysis.
\end{itemize}

\medskip

\textbf{PCE} is the most complex and computationally expensive method among the three compared. Its main advantages are:
\begin{itemize}
    \item PCE achieved excellent propagation accuracy in every scenario tested, even in the presence of large perturbations and nonlinearities. The accuracy of the approximation can be tuned according to the order of the PCE implementation, taking into account a trade-off with the computational effort.
    \item PCE is capable of approximating high-order statistical moments (i.e., skewness and kurtosis) and the entire PDF through the inexpensive generations of samples from the computed expansion at each time instant. This allows it to approximate also non-Gaussian distributions.
    \item 4th-order PCE provided results with an accuracy comparable to the Monte Carlo simulation with 10$^4$ samples, yet required approximately one order of magnitude less computational time.
\end{itemize}

\section{Conclusions}
\label{sec:conclusion}

In order to deepen the understanding of uncertainty propagation methods for close-proximity operations around small bodies, this study evaluates the performance of various techniques across different scenarios. The selected methods for comparison are Linear Covariance propagation, Unscented Transformation, Polynomial Chaos Expansion, and Monte Carlo simulation considered as the reference truth. 

Following the analysis of the dynamical environments, different scenarios are analysed. As a result, several insights are provided regarding the uncertainty propagation methods' performance and general considerations are drawn.
Linear Covariance propagation is suitable for approximating uncertainty propagation in scenarios with simple trajectories around small bodies, such as hovering arcs. Overestimates of the linear propagation can be implemented for conservative analysis in more complex scenarios.
Unscented Transformation also proves effective in the nonlinear propagation of mean and covariance in simple trajectories. Furthermore, it tends to overestimate the uncertainty distribution, making it suitable for a preliminary conservative approximation of uncertainties even in more complex and perturbed dynamical environments.
Polynomial Chaos Expansion demonstrated excellent accuracy in uncertainty propagation, even in the case of non-Gaussian distributions within complex dynamical environments. In addition, the computational time of this method is significantly lower than the one required for a Monte Carlo simulation with comparable accuracy.
Lastly, an additional noteworthy finding is that, in the case of flyby trajectories close to the asteroid surface, all methods effectively approximate uncertainty propagation within a few days around the flyby.

\section*{Declarations}

\textbf{Conflict of Interests} The authors state that there is no conflict of interest.

\newpage

\bibliography{sn-bibliography}% common bib file

\end{document}